\documentclass[pre,aps,reprint,floatfix,10pt]{revtex4-1}
\usepackage[dvips]{graphicx}
\usepackage{amsmath}
\usepackage{amssymb}
\usepackage{xcolor}
\usepackage{hyperref}

\hypersetup{
    colorlinks,
    linkcolor={red!50!black},
    citecolor={blue!50!black},
    urlcolor={blue!80!black}
}

\usepackage{float}
\usepackage{bm}

\begin{document}

%\section*{The richness of L\'evy walks}

\title{Mean squared displacement in a generalized L\'evy walk model}

\author{M. Bothe}
\affiliation{Institut f\"ur Physik, Humboldt Universit\"at zu Berlin, Newtonstra\ss e 15, 12489 Berlin, Germany}
\email{mbothe@physik.hu-berlin.de}
\author{F. Sagues}
\affiliation{Departament de Qu\'imica F\'isica, Universitat de Barcelona, 08028 Barcelona, Catalonia, Spain} 
\author{I.M. Sokolov}	
\affiliation{Institut f\"ur Physik and IRIS Adlershof, Humboldt Universit\"at zu Berlin, Newtonstra\ss e 15, 12489 Berlin, Germany}
  \email{igor.sokolov@physik.hu-berlin.de}

\begin{abstract}
L\'evy walks represent a class of stochastic models (space-time coupled continuous time random walks) with applications
ranging from the laser cooling to the description of animal motion. The initial model was intended for the description
of turbulent dispersion as given by the Richardson's law. The existence of this Richardson's regime in the original model was recently
challenged in the work by T. Albers and G. Radons, Phys. Rev. Lett. \textbf{120}, 104501 (2018): the mean squared displacement (MSD) in 
this model diverges, i.e. does not exist, in the regime, where it presumably should reproduce the Richardson's law. 
In the supplemental material to this work the authors present (but do not investigate in detail) a generalized model interpolating between 
the original one and the Drude-like models known to show no divergences. In the present work we give a detailed investigation 
of the ensemble MSD in this generalized model, show that the behavior of the MSD in this model is
the same (up to prefactiors) as in the original one in the domains where the MSD in the original model does exist, and investigate the 
conditions under which the MSD in the generalized model does exist or diverges. Both ordinary and aged situations are considered. 
\end{abstract}

\maketitle

\section{Introduction}
%Local and non-local models of L\'evy walks
A L\'evy walk, first introduced in \cite{Shlesinger}, is a space-time coupled continuous time random walk scheme, versatile enough to be applied to many different situations, ranging from 
laser cooling to animal motion (see \cite{Zaburdaev} for a comprehensive review). A genuine L\'evy walk consists of stretches of rectilinear motion. The temporal duration $t$ of a single stretch
is a random variable following the probability density function (PDF) $\psi(t)$. In an ordinary L\'evy walk the beginning of the observation coincides with the beginning of the first stretch. 
The situations when the observation starts later correspond to aged walks. Having completed a stretch, the walker changes the direction of motion at random, and a new 
stretch starts; the instant at which this happens will be called a changepoint.  The total observation time %$t$ 
after which the position of the walker is recorded does not 
have to correspond to the end of the last stretch and therefore this last stretch may stay incomplete.

A version of such a L\'evy walk commonly used for the modeling of many natural phenomena is given by a motion at a constant speed $c$, so that the displacement in the stretch $i$ is 
$\mathbf{x}_i = c \tau_i \mathbf{e}_i$ where $\mathbf{e}_i$ is a unit vector of random orientation characterizing the velocity direction during the stretch $i$ of duration $\tau_i$
. 
%First, let us concentreate at the first, ordinary, situation. 
The special case of the constant speed and fixed time in a stretch is exactly the Pearson's random walk \cite{Pearson} (in the original Pearson's question only the position at the end of 
the complete stretched was of interest). The next more complicated model is the velocity model, in which the speed during a stretch is still assumed constant but the 
duration of a stretch follows a power law $\psi(t) \simeq t^{-1-\gamma}$. The trajectory of such a walk corresponds to the collection of points in space (corresponding to the times of changepoints) 
whose spatial distribution is given by a symmetric L\'evy stable law, connected by straight lines. This fact gave the name to the whole process. 
The variants of this model where the stretches are interrupted by trapping events (rests) at changepoints arise universally when describing the particles' transport
in laminar flows consisting of eddies and jets \cite{Swinney,Swinney2,Poeschke}.

The very first work \cite{Shlesinger} however contained a discussion of a more complicated model, with the length of complete stretch depending on its duration as $|\mathbf{x}| = c t^\nu$, with 
some $\nu \geq 0$, $\nu \neq 1$ (here the constant $c$ does not have anymore the meaning of speed). It was claimed, that such a model 
is able, for example, to describe the Richardson's law \cite{Richardson} of turbulent dispersion $\langle \mathbf{x}^2 \rangle \propto t^3$ \cite{Shlesinger,Shlesinger2}.
The existence of the Richardson's regime in the original model was recently challenged in \cite{Radons}. The authors carefully investigate this original model and show that the second moment diverges 
in the domain of parameters $\gamma, \nu$ where it presumably should give rise to the Richardson's law. 

One may note that the original Levy walk model may not give the best way for mimicking transport properties in turbulent flows: 
The Richardson's law applies not to a single particle's position but to a distance between two particles in an isotropic and stationary turbulent flow. 
L\'evy-walk-like models proposed for this case, e.g. the ones of Refs. \cite{Sokolov_Drude1,Sokolov_Drude2}, differ from 
the genuine model, and rely on the isotropy, not on the spacial homogeneity of the process. The variant of the model discussed in \cite{Sokolov_Drude2} shows that in a 
particular regime of motion the process could be multiplicative: if $r_i$ is a distance between the particles corresponding to a changepoint, then it is $r_{i+1}/r_i$ and not $r_{i+1} - r_i$ 
that follows a universal distribution. Simulations of turbulent dispersion in a two-dimensional flow show the statistics of changepoints is indeed multiplicative \cite{Boffetta}.  

However the fact, unnoticed until the work \cite{Radons}, of divergence of the second moment in a model which was essentially invented to prevent divergences of moments, is extremely disturbing independently on whether this original model is a good candidate for the description of turbulant dispersion or not. 
Can this divergence be cured? Are there similar schemes which would allow for the description of the effects for which the initial scheme was devised?
How do they behave? 

Minor changes in the model bringing it closer to a Drude model for conduction in solids \cite{Drude} (as proposed in \cite{Schulz,BarKlaf}) lead to models which do 
not possess divergences in the Richardson's regime. Moreover, in the supplemental material to \cite{Radons} the authors propose, but do not discuss in detail, a model, which contains an 
additional parameter and embraces both the original L\'evy walk model, and its Drude-like variants mentioned above.  

The present work is devoted to the detailed investigation of the behavior of the mean squared displacement in this 
general model as a function of time. We consider both ordinary and aged walks, and discuss the conditions under which the mean squared displacement (MSD) does exist (i.e.
does not diverge) and investigate in detail the behavior of the MSD in the regimes when it does exist. 
In the present work we concentrate on the one-dimensional case, since it already shows all necessary richness, and since the generalization to higher dimensions is straightforward.
The notation (except for sticking to a one-dimensional model) in the present work is the same as in  \cite{Radons}.
We note that our method of solution of the problem is very straighforward, and differs from the one of Ref. \cite{Radons} footing on the Kubo-like relations.

The further structure of the present work is as follows. In Sec. \ref{Sec:Model} we introduce the generalized Levy walk model and discuss it in some detail.
General considerations of ordinary and aged L\'evy walks, and the notation used in the present work are given in Sec. \ref{Sec:Notation}.
In Sec. \ref{Sec:Ordinary} we consider the ordinary situation. Aged walks are considered in Secs. \ref{Sec:Aged} - \ref{sec:considerably} where in Sec. \ref{Sec:Aged} 
general expressions are given which are then evaluated in the two subsequent sections for special cases of aging times short and long compared to the observation time. 
The conclusions follow in Sec \ref{Sec:Conclusions}. The Appendices contain the information on the use of Tauberian theorems and details of calculations which are too lengthy to be put in the main text. 

\section{The model \label{Sec:Model}}

The generalized model discussed in the supplemental material of \cite{Radons} starts from the same distribution of the waiting times between the changepoints, $t = t_i-t_{i-1}$
which follows a power law,
\begin{equation}
 \psi(t) = \frac{\gamma}{t_0} \frac{1}{(1+ t/t_0)^{\gamma + 1}}.
 \label{PDFt}
\end{equation}
Moreover, like in \cite{Shlesinger}, the joint probability density of a stretch's length $\Delta x$ and duration $t$ in a complete stretch 
is given by 
\begin{equation}
 \psi(\Delta x,t)= \delta(|\Delta x| - c t^\nu) \psi(t).
 \label{joint}
\end{equation}

The models with the same changepoint positions still can have different behavior in dependence on the assumptions about the particle's motion
between the changepoints. 
In the original model the motion in a single stretch takes place at a constant speed, defined by the stretch duration and by its length:
\begin{equation}
 v = \frac{|\Delta x|}{t} = c t^{\nu-1}.
 \label{eq:v}
\end{equation}
The PDF of speeds can be derived from such of the waiting times and reads 
\begin{eqnarray}
p(v) &=&  \psi [t(v)] \left|d t /dv\right| \nonumber \\
&=& \frac{\gamma}{t_0 (\nu-1)} c^{\frac{-1}{\nu-1}}(1+(v/c)^{\frac{1}{\nu - 1}}/ t_0)^{-\gamma - 1} v^{\frac{2-\nu}{1-\nu}} \nonumber \\
& \simeq & c^{\frac{\gamma}{\nu-1}} t_0^{\gamma } v^{\frac{1- \gamma- \nu}{\nu -1}} \qquad \text{for } \nu \neq 1 \; .
\end{eqnarray}  
For $\nu =1$ the equation above cannot be applied (the derivative $d\tau/dv$ diverges), and one has to take $p(v) = \delta(v-c)$. For $\nu = 0$ (step length independent on the time) 
one should take $v(t)= c \delta(t)$, and this will define a ``jump first, then wait'' variant of the continuous time random walk (CTRW). 

The unpleasant property of the original model is connected with the properties of single incomplete stretches.
The velocities in such stretches in the original model are chosen according to the ``intended'' stretch duration,
and the speeds assigned to the motion over very long stetches for $\nu > 1$ are very high. 
Since the stretches whose intendent duration is long (e.g. longer than the total measurement time $t$) often stay incomplete, 
the MSD is bounded from below by the MSD in the realizations of the walk consisting only of a single stretch of duration 
not less than $t$. For $\nu > 1$ this MSD is thus $\langle x^2 \rangle \geq A(t) \int_{v(t)}^\infty v^2 t^2 p(v) dv$  
where $A(t) >0$ is the portion of single-stretch trajectories among all trajectories, and $v(t) = ct^{\nu-1}$ is the lowest speed 
in stretches which take more than $t$ to complete. For %$2-\frac{\gamma+1 + \nu}{\nu -1} \geq -1$, i.e. for 
$\nu \geq (\gamma +2)/2$ the MSD obviously diverges, as correctly stated in \cite{Radons}. 
Remarkably, such a divergence rules out the existence of the Richardson's regime (which otherwise could appear e.g. for $\nu = 3/2$ and $\gamma < 1$). 

The Drude schemes of Refs. \cite{Schulz,BarKlaf} assume that within the $i$-th stretch the law of particle's motion corresponds to $|\Delta x(t)| = c (t-t_{i-1})^\nu$,
where $t_{i-1}$ is the changepoint preceding the stretch, so that the motion proceeds not at a constant speed but accelerates or decelerates.
The speed now depends on the time in motion $t' = t-t_{i-1}$ as $v(t') = \left| \dfrac{d}{dt'} x(t')\right| = \nu c t'^{\nu-1}$.
The joint probability density of a stretch's length and time in the complete stretch 
is still given by Eq.(\ref{joint}). However, within an incomplete stretch of duration $t'$ the displacement does not depend on the intended stretch duration,
and is a deterministic function of the actual time in motion. The discussion above does not apply, and the divergences are essentially cured. 

The generalized model mixes these two situations and interpolates between the original and Drude-like model. Here the speed in a stretch does depend both on the 
intended stretch duration $\tau$ and on the actual time in motion $t'$ according to 
\begin{equation}
 v(t',\tau) = \eta c \tau^{\nu - \eta} t'^{\eta-1} , 
\end{equation}
so that in a complete stretch we always have 
\begin{equation}
 |\Delta x| = \int_0^\tau v(t',\tau) dt' = c \tau^\nu. 
\end{equation}
The displacement in an incomplete stretch $i$ is given by 
\begin{align}
 |\Delta x| = c \tau_i^{\nu - \eta} (t-t_{i-1})^\eta.
\end{align}

The original model corresponds to $\eta = 1$ and the Drude-like model to $\eta = \nu$. 
As we proceed to show, this general model shows the same asymptotic behavior of the MSD (up to prefactors) as the original model in all domains of parameters $(\gamma, \nu)$ 
where the MSD of the original model converges. Moreover we show that MSD does not diverge for $\gamma> 2(\nu- \eta)$, so that 
the Drude-like model with $\eta = \nu$ never shows divergences. These finding hold both for ordinary and aged situations. 

\section{General considerations and notation \label{Sec:Notation}}

Since the notation in our work will be somewhat more complex than in the standard problem, it is important to make it consistent and clear.
There will be two kinds of probabilities repeatedly appearing in the theory: the joint probabilities of displacements and times, and the probabilities of 
displacements conditioned on the temporal variable, and possibly on something else. 

The first ones have in 1d the dimension $[L]^{-1} [T]^{-1}$ and will be denoted by capital letters. Thus, 
let $C(x,t)$ be the joint probability density to land between $x$ and $x+dx$ in a last complete step ending between $t$ and $t+d t$ in an ordinary L\'evy walk,
observed from the instant of preparation. The observation time in an aged walk is counted from the beginning of the observation, i.e. $t=0$ corresponds to the 
time $t_a$ from the preparation of the system, and all other times are counted accordingly. Thus $F(x,t|t_a)$ is the joint probability density of the displacement in the very first step and  time after the beginning of observation at which it is completed, provided the time between the preparation of the system and the beginning of observation is $t_a$. 

\begin{figure}[h!]
\begin{center}
\includegraphics[width=85mm]{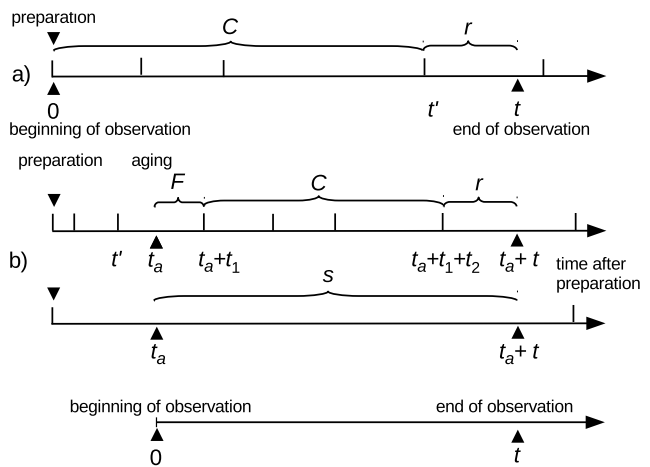}
\caption{Times and notation used in ordinary and in aged walks. a) An ordinary walk starting at preparation at $x=0$. $C$ here corresponds to a set of complete stretches
(possibly empty) and $r$ to the last, incomplete stretch starting at time $t'$. b) Aged walk. The observation starts at $t=0$, which corresponds to the time
$t_a$ after the preparation, as is depicted by the lowest time-axis in the Figure. The particle's position at the beginning of observation is declared to be the origin.
In the aged case two situations are possible: the upper scheme corresponds to the case when
the set of changepoints between the beginning of observation at $t=0$ and the final observation at $t$ is non-empty.
The time $t'$ corresponds to the last changepoint before the observation starts (it may coinside with the preparation time), the time $t_1$ to the first changepoint after the beginning of the observation, and the time
$t_2$ to the last changepoint before the observation ends (this may coinside with $t_1$). $F$ corresponds to the stretch during which the observation started.
Other notation is the same as above. The second case corresponds to a situation where there were no changepoints between $t=0$ and $t$. The two cases are mutually excluding.
\label{Not}}
\end{center}
\end{figure}

The second ones are the (conditional) probability density $p(x|t)$ of displacement at time $t$, the 
quantity we are looking for, $r(x|t)$, the conditional probability density of displacement in the last, incomplete, stretch (taking the rest of the observation time from the last changepoint), 
provided its duration is $t$, and $s(x|t,t_a)$, the conditional probability density of the position of the particle $x$ in the case when the first step after the beginning of the observation at $t=0$ is 
incomplete at the end of the observation, i.e. at time $t$. The conditional densities have the dimension $[L]^{-1}$. 
The choice of letters corresponds to $C$ -- complete, $r$ -- rest, $F$ -- first, and $s$ -- single steps. These definitions, and the times involved in the integrals below are illustrated in Fig. \ref{Not}.

Then, for the ordinary L\'evy walk we have
\begin{equation}
 p(x|t) = \int_{-\infty}^\infty dx'\int_0^t d t' C(x',t') r(x-x'|t-t').
 \label{Pord}
\end{equation}

For the aged L\'evy walk we have a more complex structure, namely 
\begin{eqnarray}
&&  p(x|t) = \int_{-\infty}^\infty dx_1 \int_{-\infty}^\infty dx_2 \int_{0}^{t} dt_1 \int_{t_1}^{t} d t_2   F(x_1,t_1|t_a) \times \nonumber \\ 
&& \;\;\; \; \times C(x_2,t_2) r(x-x_1-x_2| t-t_1-t_2) + s(x|t,t_a).
  \label{Paged}
\end{eqnarray}

In the present work we concentrate on the mean squared displacements in the corresponding walks, and note that it is possible to obtain the MSD without explicitly putting down
the PDF $p(x|t)$ in the Fourier-Laplace domain, which leads to unnecessarily complicated calculations for the aged case, which we want to treat on the same footing as an ordinary one.
To understand the structure we note that the integrals in both Eq.(\ref{Pord}) and Eq.(\ref{Paged}) have a structure of convolution in their spatial variables. Denoting
$p(k|t)$, $r(k|t)$, $F(k,t|t_a)$, $C(k,t)$, $s(k|t,t_a)$ the corresponding Fourier transforms, and noting that all 
probability densities $p,r,F,C$ and $s$ possess mirror symmetry in 1d or rotational symmetry in higher dimensions, which lets the corresponding first moments in the spatial
coordinate vanish, and assuming all of them to possess finite second spatial moment, we get that the corresponding characteristic functions have the forms 
\begin{eqnarray}
p(k|t) &=& 1- \frac{1}{2} k^2 x_2(t) + o(k^2)\\
r(k|t) &=& r_0 - \frac{1}{2} k^2(t) r_2(t) + o(k^2) \\
s(k|t,t_a) &=& s_0(t,t_a) - \frac{1}{2} k^2 s_2 (t,t_a) + o(k^2)\\ 
C(k,t) &=& C_0(t)- \frac{1}{2} k^2 C_2(t) + o(k^2)\\
F(k,t|t_a) &=& F_0(t|t_a)- \frac{1}{2} k^2 F_2(t|t_a) + o(k^2)
\end{eqnarray}  
where $x_2(t) = \langle x^2(t) \rangle$ is the total MSD, $r_2(t) = \int x^2 r(x|t) dx$ the MSD in the last, imcomplete stretch,  $s_2 (t,t_a) = \int x^2 s(x|t,t_a) dx$
the same in the case that the last stretch is a single one after the beginning of the observation, $C_2(t) = \int x^2 C(x,t) dx$ is the marginal second moment of displacement 
in a single complete step, and $F_2(t|t_a) = \int x^2 F(x,t|t_a) dx$ is the same for the first complete step after the beginning of the observation, conditioned on the aging time.
The conditions under which the second moments do exist, i.e. the corresponding coefficients in $k^2$ terms are finite, will follow from the calculations below.  

All prefactors of $k^2$ denoted by small letters have the dimension  $[L]^2$, and the ones denoted by capital letters have the dimension  $[L]^2[T]^{-1}$. 
Note that all functions except for $p$ lack normalization on unity; the zero terms of expansions are denoted $g_0 = \int g(x|t,t_a) dx \neq 1$ with 
$g$ being $r,s,Q$ or $F$. 

Passing to the Fourier-Laplace representations of $p(x|t)$ in Eq.(\ref{Pord}) and Eq.(\ref{Paged}) and taking the corresponding forms of the characteristic functions we get
\begin{equation}
 p(k,s) = C_0(s)r_0(s) - \frac{k^2}{2}\left[C_0(s) r_2(s)+C_2(s)r_0(s)\right] +o(k^2)
\end{equation}
for the ordinary process and 
\begin{eqnarray}
 p(k|s) &= &F_0(s|t_a)C_0(s)r_0(s) - \frac{k^2}{2}\left[F_0(s|t_a)C_0(s) r_2(s) \right. \nonumber \\ 
&& \left. +F_0(s|t_a)C_2(s)r_0(s)  + F_2(s|t_a)C_0(s) r_0(s)\right] \nonumber \\
&& + s_0(s,t_a) -\frac{k^2}{2}s_2(s,t_a)  +o(k^2)
 \label{aged}
\end{eqnarray}
for the aged one, where $r_i(s)$, $C_i(s)$, $F_i(s|t_a)$ and $s_i(s,t_a)$ denote the Laplace transforms of the corresponding functions 
in their $t$-variable. 
To obtain the MSD for ordinary walks we will need to get the inverse Laplace transforms of two combinations $C_0 r_2$ and $C_2r_0$, and for aged case we will need the inverse transforms of three combinations, $F_0(s)C_0(s)r_2(s)$, $F_0(s)C_2(s) r_0(s)$, $F_2(s)C_0(s) r_0(s)$ and compute $s_2$ immediately in the time domain. 
While the calculation of $C_0(s)$ and $C_2(s)$ may be performed in the Laplace domain, like for the standard model,
giving
\begin{equation}
 C(k,s) = \frac{1}{1-\psi(k,s)},
\end{equation}
other integrals have to be first evaluated in the time domain, and then transformed to the Laplace domain. 
After this the combinations $C_0 r_2$ and $C_2r_0$, as well as $F_0(s)C_0(s) r_2(s)$,  $F_0(s)C_2(s)r_0(s)$ and $F_2(s)C_0(s) r_0(s)$, have to be transformed back to time domain.

The asymptotics of Laplace transforms and the inverse Laplace transforms of functions essentially following power-laws are given by Tauberian theorems. 

These theorems give the asymptotics of a Laplace transform $f(s) = \int_0^{\infty} 
f(t) e^{-st} dt$ of a function with the large $t$ behavior $f(t) \simeq t^{\rho-1} L(t)$, 
where $L(t)$ is slowly varying, in the asymptotic regime $s \to 0$, and the corresponding inverse transforms.
%and grows at infinity slower than in an exponential manner. 
In the case $f(t)$ is non-integrable due to a divergence at $t \to \infty$, i.e. for $\rho > 0 $, Tauberian theorem can be applied directly:
\begin{equation}
 f(t) \simeq t^{\rho-1} L(t) \;\;\; \leftrightarrow \;\;\; f(s) \simeq \Gamma(\rho) s^{-\rho} L\left(\frac{1}{s}\right). \label{eqn:tauberian}
\end{equation}
In what follows we omit slowly varying functions (i.e. change them for constants, representing their limiting values $L$ at $t\to \infty$). 
By this we disregard logarithmic corrections that appear at the points of regime changes. 

If $f(t)$ is however integrable, i.e. $\rho<0$, we have to use the more general formula 
\begin{equation}
 f(s) \simeq \sum_{k=0}^{k_{max}} \frac{(-1)^k}{k!} I^{f}_k s^k +(-1)^{k_{\max}+1} L \Gamma(\rho) s^{-\rho} \; , \label{eqn:generaltauberian}
\end{equation}
where $k_{max} = \lceil -\rho-1 \rceil$ is the number of finite repeated integrals of $f(t)$, and $I^{f}_k$ is the moment integral 
\begin{equation}
I^{f}_k = \int_0^\infty t^k f(t) dt,
\end{equation}
see Appendix \ref{sec:tauberian} for detailed explanations.

\section{Ordinary walk \label{Sec:Ordinary}}

\subsection{Calculation of $\psi_0(s)$}
All properties of the ordinary walk are derived from the waiting time density $\psi(t)$. The joint probability density of a displacement in a stretch and of the time of stretch is given by
\begin{equation}
 \psi(x,t) = \frac{1}{2} \delta(|x| - ct^\nu ) \psi(t),
\end{equation}
so that the Fourier transform of this function in $x$ reads
\begin{equation}
 \psi(k,t) = \psi(t) - \frac{k^2}{2} c^2 t^{2 \nu} \psi(t) + o(k^2). 
\end{equation}
Then the Laplace transform in $t$ should be performed.  Although the exact Laplace transform of $\psi(t)$ as given by Eq.(\ref{PDFt}) is possible in quadratures, 
we are only interested in the asymptotic behavior, which can be found using the Tauberian theorem (\ref{eqn:generaltauberian}). The forms of the Laplace transform 
differ for different relations between the values of $\gamma$ and $\nu$. We are interested only in the lowest order terms
of $s$-dependence. 

\subsubsection{$ \gamma<1$}
For $0<\gamma <1$ the function $ \psi_0(s) = \psi(s)$ belongs to an integrable class, and its Laplace representation reads
\begin{equation}
 \psi_0(s) \simeq 1 + \gamma \Gamma(-\gamma) t_0^\gamma s^\gamma 
 = 1-\Gamma(1-\gamma) t_0^\gamma s^\gamma. \label{eqn:psi0GammaSmall}
\end{equation}
Keeping $t_0$ in all calculations and not putting it to unity is reasonable to be able to check the dimension of the ensuing results, especially in the aged case. 

\subsubsection{$ \gamma>1$}
The Laplace transform of $\psi(t)$ now has an additional term, due to its first moment being finite:
\begin{equation}
\psi_{0}(s) \simeq 1 - \tau s + \Gamma(1-\gamma ) t_0^\gamma s^\gamma.
\label{eqn:psi0GammaBig}
\end{equation}  
Here $\tau$ is defined as 
\begin{equation}
 \tau = \frac{\gamma}{t_0} \int_0^\infty \frac{t dt}{(1+t/t_0)^{\gamma +1}} = 
 \frac{t_0}{\gamma-1} . \label{eq:tau}
\end{equation}

\subsection{Calculation of $\psi_2(s)$}
The marginal second moment of the step distribution is given by  
\begin{align}
 \psi_2(t) = \int_{-\infty}^\infty x^2 \psi(x,t) dx = c^2 t^{2 \nu} \psi(t).
\end{align}
The expressions for $\psi_2(s)$ depend on whether $2 \nu < \gamma $ or $2 \nu > \gamma$:

\subsubsection{$2\nu<\gamma$} 
In this first case the function $\psi_2(t)$ is integrable, $\int_0^\infty \psi_2(t) dt < \infty$, and the expansion 
of its Laplace transform starts from a constant:
\begin{equation}
 \psi_2(s) \simeq \gamma c^2 t_0^{2\nu}  \int_0^\infty \frac{x^{2\nu}}{(1+x)^{\gamma+1}} dx  - \gamma \Gamma(2\nu-\gamma) c^2t_0^\gamma s^{\gamma-2\nu} \label{eqn:psi2NuSmall}
\end{equation}
where the integral is given by the dimensionless constant 
\begin{equation}
 \int_0^\infty \frac{x^{2\nu}}{(1+x)^{\gamma+1}} dx = \mathrm{B}(2\nu+1,\gamma-2\nu), \label{eqn:I1}
\end{equation}
with $\mathrm{B}(a,b)$ being the Beta-function, see Eq.(2.2.4.24) of Ref. \cite{BryPr}.  

\subsubsection{$2\nu>\gamma$} 
In the second case $\psi_2(t)$ is non-integrable, the integral $\int_0^\infty \psi_2(t) dt$ diverges, and the asymptotics of its Laplace transform read
\begin{align}
 \psi_2(s) \simeq \gamma \Gamma(2\nu-\gamma) c^2  t_0^\gamma s^{\gamma-2\nu}. \label{eqn:psi2NuBig}
\end{align}

\subsection{Calculation of $C_0(s)$ and $C_2(s)$}

For complete steps our generalization does not differ from the original L\'evy walk and we can use the general result for $C(x,t)$ in the Fourier-Laplace domain \cite{Sokolov_Klafter}:
\begin{align}
C(k,s) = \frac{1}{1 - \psi(k,s)}.
\end{align}
Expanding $C(k,s)$ for $s$ small (and fixed) and for $k \to 0$, and we get 
\begin{align}
\begin{split}
C(k,s)  &\simeq \frac{1}{1 - \psi(s) + (k^2/2) \psi_2(s) +o(k^2)}\\ &\simeq \frac{1}{1-\psi(s)} - \frac{k^2}{2}  \frac{\psi_2(s)}{[1-\psi(s)]^2} + o(k^2).
\end{split}
\end{align}
Therefore 
\begin{eqnarray}
C_0(s) &=& \frac{1}{1-\psi(s)}\\
C_2(s) &=& \frac{\psi_2(s)}{[1-\psi(s)]^2} \;,
\end{eqnarray}
into which we now have to insert our results for $\psi_0$ and $\psi_2$:

\subsubsection{$\gamma<1$ and $2\nu<\gamma$ }
Here we find by using equations (\ref{eqn:psi0GammaSmall}) and (\ref{eqn:psi2NuSmall}) :
\begin{eqnarray}
 C_0(s) &\simeq & \frac{1}{\Gamma(1-\gamma)} t_0^{-\gamma} s^{-\gamma}, \label{eq:Co}\\
 C_2(s) &\simeq & \gamma   \frac{ \mathrm{B}(2\nu+1,\gamma-2\nu)}{\Gamma^2(1-\gamma)} c^2 t_0^{2\nu - 2\gamma} s^{-2 \gamma} .
\end{eqnarray}

\subsubsection{$\gamma<1 \text{ and } 2\nu>\gamma$: }
In this case we obtain for $C_2$:
\begin{equation}
  C_2(s) \simeq  \gamma   \frac{\Gamma(2\nu-\gamma)}{\Gamma^2(1-\gamma)} c^2 t_0^{-\gamma} s^{-2\nu-\gamma},
\end{equation}
while $C_0$ is the same as in Eq.(\ref{eq:Co}) as long as $\gamma<1$.

\subsubsection{$\gamma>1$ and  $2\nu<\gamma$}
Now the first moment of $\psi(t)$ is finite, therefore 
\begin{eqnarray}
 C_0(s) &\simeq &  \frac{1}{\tau s}  \\
 C_2(s) &\simeq & \gamma \mathrm{B}(2\nu+1,\gamma-2\nu) c^2 \frac{t_0^{2\nu}}{\tau^2}  s^{-2}
\end{eqnarray}

\subsubsection{$\gamma>1$ and $2\nu>\gamma$: }
Again $C_0(s)$ doesn't change, and for $C_2$ we obtain
\begin{equation}
C_2 \simeq  \gamma \Gamma(2\nu-\gamma)  c^2 \frac{t_0^\gamma}{\tau^2}   s^{\gamma-2\nu -2} .
\end{equation}

\subsection{Calculation of  $r_0(s)$ and $r_2(s)$}
The function $r(x|t)$ gives the distribution of
the corresponding displacements conditioned on the fact that the total duration of a stretch is longer than $t$:
\begin{equation}
 r(x|t) =  \int_t^\infty \frac{1}{2} \delta(|x| - c t^{\eta}t'^{\nu-\eta}) \psi(t') dt' \; .
\end{equation}
This is correctly normalized to the overall probability to stay within a single stretch for a time longer than $t$ 
\begin{equation}
 \int r(x|t)  dx = \int_t^\infty \psi(t') dt'.
\end{equation}
Expanding the Fourier transform of $r(x,t)$ for small $k$,
\begin{eqnarray}
r(k|t) &=& \int_{-\infty}^{\infty} dx \; e^{ikx} \int_t^\infty \frac{1}{2} \delta(|x| - c t^{\eta}t'^{\nu-\eta}) \psi(t') dt' \nonumber \\
&=& r_0(t) - \frac{k^2}{2} r_2(t) + o(k^2), 
\end{eqnarray}
we find the marginal moments
\begin{eqnarray}
r_0(t) &=& \frac{1}{(1+t/t_0)^{\gamma}} \\
r_2(t) &=& \gamma c^2 \frac{1 }{t_0} t^{2\eta} \int_t^\infty \frac{t'^{2(\nu-\eta)}}{(1+t'/t_0)^{\gamma+1}} dt' \; , 
\end{eqnarray}
whose Laplace transforms depend on the relationship between $\gamma$, $\nu$ and $\eta$: 

\subsubsection{$\gamma<1 $ }
In this case $r_0$ is non-integrable and we find
\begin{equation}
  r_0(s) \simeq  \Gamma(1-\gamma) t_0^\gamma s^{\gamma-1}.
\end{equation}
For $r_2$  we can use the asymptotic form of $\psi(t)$ , since we are interested in large $t$:
\begin{equation}
r_2(t) \simeq \gamma c^2 t_0^\gamma t^{2\eta} \int_t^\infty \tau^{2(\nu-\eta)-1-\gamma}  d\tau   \; , 
\end{equation}
which is only finite for $\gamma > 2(\nu-\eta)$ and diverges otherwise, meaning that no MSD exists. In the rest of 
the paper we will concentrate on the case when the second moment converges, where we find
\begin{equation}
r_{2}(t)  \simeq \gamma\frac{1} {2(\nu-\eta)-\gamma} c^2 t_0^\gamma t^{2\nu -\gamma}  .
\end{equation}
Since this expression is always in the non-integrable class we obtain by using Tauberian theorem (\ref{eqn:tauberian})
\begin{equation}
r_2(s) \simeq \gamma  \frac{ \Gamma(2\nu+1-\gamma)}{2(\nu-\eta)-\gamma}c^2 t_0^\gamma s^{\gamma - 2\nu-1}.
\end{equation}

\subsubsection{$\gamma>1$ and $2\nu>\gamma-1$}
For $\gamma>1$ $r_0(t)$ becomes integrable, therefore its Laplace transform reads
\begin{equation}
 r_0 \simeq \tau - \Gamma(1-\gamma) t_0^\gamma s^{\gamma-1} .
\end{equation}
The function $r_2(t)$ is still non-integrable and therefore identical to the previous case.

\subsubsection{$\gamma>1$ and $2\nu<\gamma-1$}
In this case $r_0$ does not change but $r_2(s)$ is now integrable and therefore its transform behaves as
\begin{equation}
r_2(s) \simeq I^{r_2}_0 - \gamma \frac{\Gamma(2\nu+1-\gamma)}{2(\nu-\eta)-\gamma} c^2 t_0^\gamma s^{\gamma - 2\nu-1 } \; .
\end{equation}
We evaluate the definite integral $I^{r_2}_0$ by manipulating the area of integration:
\begin{eqnarray}
 I^{r_2}_0 &=& c^2\int_0^\infty dt  t^{2\eta} \int_t^\infty dt' \psi(t') (t')^{2(\nu-\eta)} \\
  &=& c^2\frac{1}{2\eta+1} \int_0^\infty \psi(t') t'^{2\nu+1} dt' \nonumber \\ 
 &=& \gamma\frac{1}{2\eta+1} \int_0^\infty \frac{x^{2\nu+1}dx}{(1+x)^{\gamma+1}}c^2 t_0^{2\nu+1}  \nonumber \\
&=& \gamma  \frac{\mathrm{B}(2\nu +2, \gamma - 2\nu -1)}{2\eta+1}  c^2 t_0^{2\nu+1} , 
\end{eqnarray}
meaning the $r_2$ is constant in the leading order. Therefore we find for $\gamma>1$ and $2\nu<\gamma-1$

\begin{eqnarray}
r_2 &\simeq & \gamma \frac{\mathrm{B}(2\nu +2, \gamma - 2\nu -1)}{2\eta+1}  c^2 t_0^{2\nu+1}   \nonumber \\
 & &  -\gamma \frac{\Gamma(2\nu+1-\gamma)} {2(\nu-\eta)-\gamma} c^2 t_0^\gamma  s^{\gamma - 2\nu-1} .
\end{eqnarray}

The results so far are summarized in Table \ref{tab:Cr}.

\begin{widetext}
\begin{center}
\begin{table}[htb]
 \begin{tabular}{||l|l|l||}
 \hline \hline
 \rule[-4mm]{0cm}{1cm} $\gamma<1$ & 
 $C_0(s) \simeq \frac{1}{\Gamma(1-\gamma)}t_0^{-\gamma} s^{-\gamma}$ & 
 $C_2(s) \simeq \left\{
\begin{array}{ll}
\gamma \frac{\mathrm{B}(2\nu+1,\gamma-2\nu)}{\Gamma^2(1-\gamma)} c^2 t_0^{2\nu - 2\gamma} s^{-2 \gamma} &\mbox{for } 2\nu < \gamma \\
 \gamma \frac{\Gamma(2\nu-\gamma)}{\Gamma^2(1-\gamma)} c^2 t_0^{-\gamma} s^{-2\nu-\gamma} &\mbox{for } 2\nu > \gamma
\end{array} \right. $ \\ \hline
\rule[-4mm]{0cm}{1cm}  $\gamma<1$ &$r_0(s) \simeq \Gamma(1-\gamma)t_0^\gamma s^{\gamma-1}$ & $r_2(s)  \simeq \gamma \frac{\Gamma(2\nu+1-\gamma)}{2(\nu-\eta)-\gamma}c^2 t_0^\gamma s^{\gamma-1-2\nu}$ \\ \hline
\rule[-4mm]{0cm}{1cm}   $\gamma>1$ & $C_0(s) \simeq   \frac{1}{\tau s}$ & 
 $C_2(s) \simeq \left\{ \begin{array}{ll}
 \gamma \mathrm{B}(2\nu+1,\gamma-2\nu)c^2 \frac{t_0^{2\nu}}{\tau^2}  s^{-2} & \mbox{for } 2\nu < \gamma \\
\gamma \Gamma(2\nu-\gamma)c^2\frac{t_0^\gamma}{\tau^2} s^{\gamma-2\nu -2}  & \mbox{for }  2\nu > \gamma
\end{array} \right. $\\ \hline
\rule[-4mm]{0cm}{1cm} $\gamma>1$ &$r_0(s) \simeq\ \tau$ 
 & $r_2(s) \simeq \left\{ \begin{array}{ll}
\gamma\frac{\mathrm{B}(2\nu +2, \gamma - 2\nu -1)}{2\eta+1}  c^2 t_0^{2\nu+1}  & \mbox{for } 2\nu < \gamma -1 \\
\gamma \frac{\Gamma(2\nu+1-\gamma)}{2(\nu-\eta)-\gamma}  c^2 t_0^\gamma   s^{\gamma - 2\nu-1}& \mbox{for } 2\nu > \gamma -1 
\end{array} \right.$ \\ \hline \hline
\end{tabular}
\caption{Leading terms of the marginal moments of $C$  and $r$ in the Laplace domain for different parameter ranges. \label{tab:Cr}}
\end{table}
\end{center}
\end{widetext}

\subsection{Mean squared displacement}
With these results we can now compute the MSD via the formula
\begin{equation}
\langle x^2(s) \rangle = C_0(s) r_2(s) + C_2(s) r_0(s) \; .
\end{equation}

\subsubsection{$\gamma<1$ and $2\nu < \gamma$: }
In the case  $2\nu < \gamma $ we have
\begin{equation}
\begin{split}
\langle x^2(s) \rangle  \simeq &  \gamma  \left[\frac{\Gamma(2\nu +1-\gamma)}{\Gamma(1-\gamma)(2(\nu-\eta)-\gamma)}c^2 s^{-2\nu -1} \right. \\
& \left. + \frac{\mathrm{B}(2\nu+1,\gamma-2\nu)}{\Gamma(1-\gamma)}c^2 t_0^{2\nu-\gamma} s^{-\gamma -1} \right] , 
\end{split}
\end{equation}
which translates to
\begin{equation}
\begin{split}
 \langle x^2(t) \rangle \simeq \gamma  & \left[\frac{\Gamma(2\nu +1-\gamma)}{\Gamma(1-\gamma)(2(\nu-\eta)-\gamma)\Gamma(2\nu + 1)} c^2 t^{2\nu} \right. \\ 
 & \left.  + \frac{\mathrm{B}(2\nu+1,\gamma-2\nu)}{\Gamma(1-\gamma)\Gamma(1+\gamma)} c^2 t_0^{2\nu-\gamma} t^{\gamma} \right].
 \end{split}
\end{equation}
This is dominated by the second term since $2\nu < \gamma$, leading to  $\langle x^2(t) \rangle \propto t^\gamma$ in the case. 
This means that for $\gamma < 1$ and $2\nu < \gamma$ the behavior of the walk merges with the one of a CTRW
with a fixed step length.

\subsubsection{$\gamma<1$ and  $2\nu > \gamma$ }
In this parameter regime we obtain
\begin{eqnarray}
 \langle x^2(s) &\simeq & \gamma  \nonumber \\
 && \times \left[\frac{\Gamma(2\nu+1-\gamma)}{\Gamma(1-\gamma)(2(\nu-\eta)-\gamma)} +  \frac{\Gamma(2\nu-\gamma)}{\Gamma(1-\gamma)} \right] c^2 s^{-2\nu-1} \nonumber \\
 &= & \gamma  \frac{\Gamma(2\nu -\gamma)}{\Gamma(1-\gamma)} \frac{4\nu-2\eta -2 \gamma}{2(\nu-\eta)-\gamma}c^2  s^{-2\nu-1}
\end{eqnarray}
therefore we find in the time domain
\begin{equation}
 \langle x^2(t) \rangle \simeq \gamma \frac{\Gamma(2\nu -\gamma)} {\Gamma(2\nu+1) \Gamma(1-\gamma)} \frac{4\nu-2\eta -2 \gamma}{2(\nu-\eta)-\gamma}  c^2  t^{2\nu} .
\end{equation}

\subsubsection{$\gamma>1$ and $2\nu < \gamma-1$}
In this case the MSD reads
\begin{eqnarray}
\langle x^2(s) \rangle &=& C_0(s) r_2(s) + C_2(s)  r_0(s) \nonumber \\
& \simeq & \gamma \frac{\mathrm{B}(2\nu +2, \gamma - 2\nu -1)}{2\eta+1} c^2 \frac{t_0^{2\nu+1}}{\tau}\frac{1} {s}    \\
& & + \gamma \mathrm{B}(2\nu+1,\gamma-2\nu) c^2 \frac{t_0^{2\nu}}{\tau}  s^{-2} \nonumber ,
\end{eqnarray}
which is dominated by the second term. Therefore we find in the time domain in leading order:
\begin{equation}
\langle x^2(t) \rangle \simeq \gamma \mathrm{B}(2\nu+1,\gamma-2\nu) c^2 \frac{ t_0^{2\nu}}{\tau}  t.
\end{equation}

\subsubsection{$\gamma>1 \text{ and }  \gamma-1 <2\nu < \gamma$}
Compared to the previous case only $r_2$ changes, therefore
\begin{eqnarray}
\langle x^2(s) \rangle &\simeq & \gamma \frac{\Gamma(2\nu+1-\gamma)}{2(\nu-\eta)-\gamma} c^2\frac{t_0^{\gamma}}{\tau}  s^{\gamma-2-2\nu} \nonumber \\
&& + \gamma \mathrm{B}(2\nu+1,\gamma-2\nu) c^2\frac{t_0^{2\nu}}{\tau}  s^{-2} .
\end{eqnarray}
Since $\gamma-2\nu>0$ the term quadratic in $s$ is again dominant, and the asymptotic behavior in the time domain is identical to the previous case:
\begin{equation}
\langle x^2(t) \rangle \simeq   \gamma \mathrm{B}(2\nu+1,\gamma-2\nu) c^2
\frac{t_0^{2\nu}}{\tau}  t.
\end{equation}

\subsubsection{$\gamma>1$ and $2\nu > \gamma$}
Now $C_2$ is different, giving us 

\begin{eqnarray}
\langle x^2(s) \rangle &\simeq & \gamma \frac{\Gamma(2\nu+1-\gamma)}{2(\nu-\eta)-\gamma} c^2 \frac{t_0^{\gamma}}{\tau}  s^{\gamma-2\nu-2} \nonumber\\
& & + \gamma \Gamma(2\nu-\gamma) c^2\frac {t_0^{\gamma}}{\tau}  s^{\gamma-2\nu-2}. 
\end{eqnarray}
This results in the time dependence 
\begin{equation}
\langle x^2(t) \rangle \simeq \gamma  \Gamma(2\nu-\gamma) \frac{4\nu -2 \eta -2\gamma}{2(\nu-\eta) -\gamma}c^2  \frac{t_0^{\gamma}}{\tau } t^{2\nu+1-\gamma}.
\end{equation}

The results for the ordinary walk under the assumption that the convergence condition $\gamma > 2(\nu-\eta)$ is satisfied can be summarized as follows: 
\begin{equation}
\langle x^2(t) \rangle \propto \left\{
  \begin{array}{lll}
  t^{\gamma} & \mathrm{for} & \gamma < 1, \; 2\nu < \gamma \\
t^{2\nu} & \mathrm{for} & \gamma < 1, \; 2\nu > \gamma \\
t & \mathrm{for} & \gamma > 1, \; 2\nu < \gamma \\
t^{2\nu+1- \gamma  } &\mathrm{for} &  \gamma > 1, \;2\nu > \gamma .
  \end{array}
  \right.
\label{balpoint}
\end{equation}
Thus, in the whole domain of $\gamma$ there are four regimes with crossovers at $\gamma = 1$ and at $2\nu = \gamma$:
\begin{itemize}
 \item For $2\nu < \gamma$ one has $\langle x^2(t) \rangle \propto t^{\gamma}$ for $\gamma <1$ crossing over to a faster growth $\langle x^2(t) \rangle \propto t$ for $\gamma > 1$
 \item For $2\nu > \gamma$ one has universally $\langle x^2(t) \rangle \propto t^{2\nu}$ for $\gamma <1$ crossing over to a slower growth $\langle x^2(t) \rangle \propto t^{2\nu+1-\gamma}$ for $\gamma > 1$.
\end{itemize}
Note that in an ordinary walk the Richardson regime is possible, and is achieved for $\nu = 3/2$ for $\gamma <1$, and for $\nu = (\gamma + 2)/2$ for $\gamma > 1$, provided $\gamma > 2(\nu - \eta)$. 
The last restriction corresponds to $\eta > (3-\gamma)/2$ for $\gamma <1$, and to $\eta >1$ for $\gamma > 1$. The original model with $\eta =1$ indeed does not possess a Richarson regime. 

\section{Aged walk: General expressions \label{Sec:Aged}}

We now consider the functions $F$ and $s$ which are specific for aged walks. The general expression for $F$ reads: 
\begin{eqnarray}
&& F(x,t|t_a)= \int_0^{t_a} dt'  \psi(t_a+t-t') k(t') \times \\
&& \;\; \times \delta \left\{x-c[(t_a+t-t')^\nu-(t_a+t-t')^{\nu-\eta} (t_a-t')^\eta]\right\} , \nonumber
\end{eqnarray}
where $k(t)=C_{0}(t)$ is the time-dependent rate of steps. Note that the argument of the $\delta$-function is shifted, 
due to the fact that the distance from the origin $x$ is set to zero at the start of the measurement.
The marginal normalization of $F(x,t|t_a)$ is
\begin{eqnarray}
F_0(t|t_a) &=& \int F(x,t|t_a) dx =\int_0^{t_a} \psi(t_a+t-t') k(t') dt' \nonumber \\
 &=& \psi_1(t|t_a),
\end{eqnarray}
where $\psi_1(t|t_a)$ is the forward waiting time PDF known from the theory of continuous time random walk (CTRW) \cite{Sokolov_Klafter}. The marginal second moment of $F$ reads:
\begin{eqnarray}
F_2(t|t_a) &=& \int_0^{t_a} dt' c^2 \psi(t_a+t-t') k(t') \times \label{eqn:F2} \\ 
&& \times [(t_a+t-t')^\nu-(t_a+t-t)^{\nu-\eta} (t_a-t')^\eta]^2 . \nonumber
\end{eqnarray}
Additionally we have to consider the term $s(x|t,t_a)$, which describes the case that both the aging time and the observation time belong to the same stretch:
\begin{eqnarray}
&& s(x|t,t_a) = \int_0^{t_a} dt' k(t') \int^{\infty}_{t_a+t-t'} dt'' \psi(t'') \times \\
&& \;\; \times \delta \left\{x-c[(t'')^{\nu-\eta}(t_a+t-t')^\eta-(t'')^{\nu-\eta}(t_a-t')^{\eta}]\right\} , \nonumber 
\end{eqnarray}
where the inner integral gives the probability that no renewal took place during the time interval between $t'$ and $t_a+t$. The normalization of this function, giving $s_0$, 
\begin{equation}
 s_0(t,t_a) = \int_0^{t_a} \Psi(t_a+t-t') k(t') dt' 
\end{equation}
where $\Psi(t) = \int_t^\infty \psi(t') dt'$ is the survival probability,
is not necessary for what follows, and will not be calculated. The second moment is given by
\begin{eqnarray}
&& s_2(t,t_a) =c^2 \int_0^{t_a}  k(t') \int^{\infty}_{t_a+t-t'}  \psi(t'') \times \label{eqn:s2}\\ 
&& \;\; \times [(t'')^{\nu-\eta}(t_a+t-t')^\eta-(t'')^{\nu-\eta}(t_a-t')^{\eta}]^2   dt'' dt'. \nonumber 
\end{eqnarray}
We note that the form of the integrals involved in $F_2$ and $s_2$ is very similar. 
In the following calculation we differentiate between two time regimes: The case of short aging times $t>>t_a>> t_0$ and the case of long aging times $t_a>>t>>t_0$, 
which will be discussed separately in the two following sections. 

\section{Short aging times $t>> t_a$}

\subsection{Calculation of $F_0$ and $F_2$}

We are interested in the Laplace transforms of the marginal moments of $F_0$ and $F_2$. Since both of them depend on $k(t)=C_0(t)$ 
whose form we found to be dependent on whether $\gamma<1$ or $\gamma>1$, we have to distinguish between these cases.

\subsubsection{$\gamma<1$}
The function $F_0(t|t_a)$ is equal to the forward waiting time PDF $\psi_1(t|t_a)$. For $\gamma<1$
it is given by the following expression:
\begin{equation}
 F_0(t|t_a) = \psi_1(t|t_a) = \frac{\sin \pi \gamma}{\pi} \left( \frac{t_a}{t}\right)^\gamma \frac{1}{t+t_a}. \label{eqn:psi1}
\end{equation}
The expression is normalized to unity, and therefore in the Laplace domain the leading term in $F_0$ will be 1. 

For  $t>>t_a$ the expression in the square brackets in Eq.(\ref{eqn:F2}) can be approximated by $t^{2\nu}$ (since $\nu > 0$) and for $F_2$ we find in this limit
\begin{equation}
 F_2(t|t_a) \simeq  c^2 t^{2\nu} \int_0^{t_a} \psi(t_a+t-t') k(t') dt'.
\end{equation}
The integral can again be expressed through the forward waiting time $\psi_1(t|t_a)$. We take the asymptotics of $\psi_1$ for $t$ large, so that 
\begin{equation}
F_2(t|t_a) \simeq \frac{\sin \pi \gamma}{\pi} c^2  t_a^\gamma t^{2\nu-\gamma -1}  .  
\end{equation}
Here again two situations arise depending on the integrability: 

\subsubsection{$\gamma <1$ and  $2 \nu > \gamma$}
In this case $F_2$ is non-integrable, so that in the Laplace domain
\begin{equation}
 F_2(s|t_a) \simeq \Gamma(2\nu-\gamma) \frac{\sin \pi \gamma}{\pi} c^2 t_a^\gamma s^{\gamma - 2\nu}.
\end{equation}

\subsubsection{$\gamma <1$  and $2\nu < \gamma $} 
Now $F_2$ is integrable, and the lowest order in its Laplace transform tends to a constant:
\begin{equation}
 F_2(s|t_a) \simeq \frac{\sin \pi \gamma}{\pi}  c^2 t_a^\gamma \int_0^\infty \frac{ t^{2\nu-\gamma}}{t+t_a} dt.
\end{equation}
The corresponding integral is given by
\begin{equation}
 \int_0^\infty \frac{t^{2\nu-\gamma}}{t+t_a} dt = \frac{\pi }{\sin (\pi(2\nu +1- \gamma))}t_a^{2\nu - \gamma} ,
\end{equation}
see Eq.(2.2.5.25) of Ref. \cite{BryPr}, so that
\begin{equation}
 F_2(s|t_a) \simeq   \frac{\sin \pi \gamma}{\sin (\pi(2\nu +1- \gamma))}c^2 t_a^{2\nu}.
\end{equation}

\subsubsection{$\gamma > 1$}
Now we consider the case $\gamma > 1$. From the previous section we know that $C_0(t)=\frac{1}{\tau s}$, therefore
\begin{equation}
k(s) = \frac{1}{\tau} . \label{eqn:kGammaLarge}
\end{equation}
With this we can rewrite $F_0$ as 
\begin{equation}
 F_0(t|t_a) = \frac{1}{\tau} \int_0^{t_a} \psi(t+y)dy.
\end{equation}
For $t \to \infty$ it decays as $t^{-\gamma}$ and therefore is of integrable type. To find its lowest order (constant) term we note that
\begin{eqnarray}
 F_0 &=& \frac{1}{\tau} \int_0^\infty dt \int_0^{t_a} \psi(t+y)dy = \frac{1}{\tau} \int_0^{t_a} dy \int_0^\infty dt \psi(t+y) \nonumber \\
 &=& \frac{1}{\tau} \int_0^{t_a} dy \int_y^\infty dt \psi(t) = \frac{1}{\tau} \int^{t_a}_0 \Psi(y) dy,
\end{eqnarray}
which tends to unity since $t_a \gg t_0$ and since $\int_0^\infty \Psi(t') dt' = \tau$. 
The term $F_2$ for $t \gg t_a$ can again be 
evaluated by approximating the expression in square brackets in Eq.(\ref{eqn:F2}) by $t^\nu$: 
\begin{eqnarray}
 F_2(t|t_a) &\simeq & c^2  \frac{1}{\tau} t^{2\nu} \int_0^{t_a} \psi(t+y)dy \\
 &=& c^2 \frac{ t_0^\gamma }{\tau} [(t+t_0)^{-\gamma}-(t+t_a+t_0)^{-\gamma}]t^{2\nu}. \nonumber
 \label{F2lagrer1}
\end{eqnarray}
Since the power-law asymptotics of the expression in square brackets is $t^{-\gamma -1}$ the whole expression
\begin{equation}
 F_2(t|t_a) \simeq \gamma c^2  \frac{t_0^\gamma }{\tau}t_a t^{2\nu-\gamma -1}
\end{equation}
is of the non-integrable type for $2\nu > \gamma$ and of integrable type for $2 \nu < \gamma$.

\subsubsection{$\gamma>1$ and $2\nu > \gamma$} 
In this first case $F_2(t|t_a)$ is non-integrable, therefore the Laplace transforms is 
\begin{equation}
 F_2(s|t_a) \simeq \gamma\Gamma (2\nu-\gamma) c^2\frac{t_0^\gamma}{\tau}t_a s^{\gamma - 2\nu}.
\end{equation}

\subsubsection{$\gamma>1$ and $2\nu < \gamma$}
In the second case the Laplace transform of the expression tends to a constant. To evaluate this we put down 
\begin{equation}
 F_2 \simeq  \gamma c^2 \frac{t_0^\gamma}{\tau} \int_0^\infty dt \; t^{2\nu} \int_0^{t_a} \frac{1}{(t+t_0+y)^{\gamma+1}} dy \; ,
\end{equation}
and interchange the sequence of integrations:
\begin{eqnarray}
 F_2 &=& \gamma c^2 \frac{ t_0^\gamma}{\tau} \int_0^{t_a} dy \int_0^\infty \frac{t^{2\nu}}{(t+t_0+y)^{\gamma+1}} dt \nonumber \\
 &=& \gamma \mathrm{B}(2\nu+1,\gamma - 2\nu) c^2 \frac{ t_0^\gamma}{\tau}  \int_0^{t_a} (t_0+y)^{2\nu -\gamma} dy \nonumber \\
 &=& \gamma  \frac{\mathrm{B}(2\nu+1,\gamma - 2\nu) }{2 \nu +1- \gamma } c^2 \frac{t_0^\gamma}{\tau} \times \\
 && \;\;\; \times [(t_0+t_a)^{2\nu +1- \gamma }-t_0^{2\nu+1 - \gamma }]. \nonumber
\end{eqnarray}
(in the transition to the second line the Eq.(2.2.5.24) of Ref. \cite{BryPr} is used). The corresponding expression is dominated by the first or by the second term in the square 
brackets, depending on whether $2\nu > \gamma -1$ or $2\nu < \gamma -1$. \\
We summarize our results for $\gamma > 1$ in the following formula:
\begin{equation}
 F_2 = \left\{
 \begin{array}{ll}
   \gamma \frac{\mathrm{B}(2\nu+1,\gamma - 2\nu)}{2 \nu - \gamma +1} c^2 \frac{  t_0^{2\nu +1}}{\tau}  & \mbox{for } 2\nu < \gamma -1 \\
  \gamma \frac{\mathrm{B}(2\nu+1,\gamma - 2\nu)}{2 \nu - \gamma +1} c^2 \frac{ t_0^\gamma}{\tau} t_a^{2\nu +1- \gamma }  & \mbox{for } \gamma-1 < 2\nu < \gamma \\ 
 \gamma \Gamma(2\nu-\gamma) c^2 \frac{t_0^\gamma}{\tau} t_a s^{\gamma - 2\nu} & \mbox{for } 2\nu > \gamma 
 \end{array}
 \right.
\end{equation}

\subsection{Calculation of $s_2$}
We can calculate the second marginal moment of the single step PDF $s_{2}(t,t_a)$ directly in the time domain. For this we need the stepping rate $k(t)=C_0(t)$, whose behavior depends on whether $\gamma>1$ or $\gamma<1$.

\subsubsection{$\gamma<1$}
In this case we find by inverse transform of $C_0$ from table \ref{tab:Cr}:
\begin{equation}
 k(t) = \frac{1}{\Gamma(\gamma) \Gamma(1-\gamma)} t_0^{-\gamma} t^{\gamma-1} = \frac{\sin \pi \gamma}{\pi} t_0^{-\gamma} t^{\gamma-1}. \label{eqn:kGammaSmall}
\end{equation}
Inserting this result into equation (\ref{eqn:s2}) we find
\begin{eqnarray}
s_2(t,t_a) &=& \gamma\frac{\sin(\pi \gamma)}{\pi}  c^2 \times \\
&& \;\;\; \times \int_0^{t_a} \int^{\infty}_{t_a+t-t'} [(t_a+t-t')^\eta-(t_a-t')^{\eta}]^2 \nonumber\\
&& \;\;\; \times   (t')^{\gamma-1}  (t'')^{2(\nu-\eta)} \frac{1}{(t_0+t'')^{\gamma+1}} dt'' dt' . \nonumber
\end{eqnarray}
Just like $r_2$ in the ordinary case, this integral only converges for $\gamma>2(\nu-\eta)$, meaning that $\eta$ again governs the existence of the second moment. In the limit $t_0<< t''$ we can write:
\begin{eqnarray}
s_2(t,t_a) &\simeq&  \gamma\frac{\sin(\pi \gamma)}{\pi} \frac{ 1 }{\gamma-2(\nu-\eta)} c^2 \int_0^{t_a} (t_a+t-t')^{2(\nu-\eta)-\gamma}   \nonumber \\
&& \times [(t_a+t-t')^\eta-(t_a-t')^{\eta}]^2 (t')^{\gamma-1}  dt' .
\end{eqnarray}
The expression in square brackets is again approximated by $t^{2\eta}$ and $(t_a+t-t')$ by its value at $t$, so that
\begin{equation}
 s_2(t,t_a) \propto c^2 t_a^\gamma t^{2\nu -\gamma}. 
\end{equation}

\subsubsection{$\gamma>1$}
In this regime we have $k(t)=\frac{1}{\tau}$ again. Substituting this into equation (\ref{eqn:s2}) and  approximating the term in square brackets again results in
\begin{equation}
s_2(t,t_a) \simeq \gamma \frac{1}{\gamma-2(\nu-\eta)}c^2 \frac{t_0^{\gamma}}{\tau} t^{2\eta} \int_{0}^{t_a} (t_a+t-t')^{2(\nu-\eta)-\gamma} dt'  ,
\end{equation}
which gives us
\begin{equation}
s_2(t,t_a) \simeq \gamma \frac{1}{(\gamma-2(\nu-\eta))(\gamma-2(\nu-\eta)-1)} c^2\frac{t_0^{\gamma}}{\tau} t_a t^{2\nu-\gamma} \; .
\end{equation}
The results so far are summarized in Table \ref{tab:sFweakAging}, where we used the fact that $\tau \propto t_0$, see Eq.(\ref{eq:tau}). 

\begin{widetext}
\begin{center}
\begin{table}[h!]
 \begin{tabular}{||l|l|l||}
 \hline \hline
\rule[-4mm]{0cm}{1cm}   $\gamma<1$ & $F_0(s|t_a) \simeq 1$ & $F_2(s|t_a) \propto \left\{
\begin{array}{ll}
c^2 t_a^{2\nu}  & \mbox{for } 2\nu < \gamma \\
c^2 t_a^\gamma s^{\gamma-2\nu} & \mbox{for } 2\nu > \gamma
\end{array} \right. $\\ \hline
\rule[-4mm]{0cm}{1cm} $\gamma<1$ & $  $ & $s_2(t|t_a) \propto c^2 t_a^\gamma t^{2\nu -\gamma} $ \\ \hline
\rule[-4mm]{0cm}{1cm} $\gamma>1$ & $F_0(s|t_a) \simeq 1$ & $F_2(s|t_a) \propto \left\{
 \begin{array}{ll}
c^2  t_0^{2\nu} & \mbox{for } 2\nu < \gamma  -1     \\
  c^2  t_0^{\gamma-1} t_a^{2\nu +1- \gamma } & \mbox{for } \gamma-1 < 2\nu < \gamma \\
 c^2 t_0^{\gamma-1} t_a s^{\gamma - 2\nu} & \mbox{for } 2\nu > \gamma
 \end{array}
 \right. $ \\ \hline
\rule[-4mm]{0cm}{1cm} $\gamma>1$& & $s_2(t,t_a) \simeq c^2 t_0^{\gamma-1}  t_a t^{2\nu-\gamma}$ \\ \hline \hline
\end{tabular}
\caption{Results for $F_0$ and $F_2$ in the Laplace domain as well as $s_2$ in the time domain for different parameter ranges in the case of weak aging $t>>t_a$. Dimensionless prefactors are omitted. \label{tab:sFweakAging}}
\end{table}
\end{center}
\end{widetext}

\subsection{Mean squared displacement}

We are now ready to calculate the MSD in the weakly aged case. Recall our earlier result
\begin{eqnarray}
\langle x^2 \rangle (s|t_a) &=&  F_0(s|t_a) C_0(s) r_2(s) + F_0(s|t_a) C_2(s) r_0(s)  \nonumber \\
&& \;\; + F_2(s|t_a) C_0(s) r_0(s) + s(s|t_a) .
\end{eqnarray}
We can now write down the first three terms in the Laplace domain using the results from the tables \ref{tab:Cr} and \ref{tab:sFweakAging}, and transform them back into the time domain.
The last term in the time domain is already known. The calculation results in different asymptotics depending on $\gamma$.

\subsubsection{$\gamma<1$}
In this regime the terms $F_0C_0 r_2 $ and $ F_0C_2 r_0$ reproduce the result for the non-aged walks. The term $F_2C_0 r_0 (s)  \propto t_a^\gamma s^{\gamma-2\nu -1}$
translates into $F_2C_0 r_0 (t) \propto c^2 t^{2\nu}(t_a/t)^\gamma$, and is subdominant for $t \gg t_a$ for $2\nu > \gamma$. For $2\nu < \gamma$ this term tends to
$const \cdot  c^2 t_a^{2\nu}s^{-1}$, i.e. is a constant proportional to $c^2 t_a^{2\nu}$ in the time domain, and is again subdominant with respect to the previous ones.
The term $s_2$ has the same asymptotics as the previous one in the first case, $s_2 \propto c^2 t^{2\nu}(t_a/t)^\gamma$ and therefore is also subdominant. 

\subsubsection{$\gamma>1$}
For this regime the contributions $F_0C_0 r_2 $ and $F_0C_2 r_0 $ give the same behavior as in the non-aged case, $\propto t^{2\nu +1- \gamma }$ for $2\nu > \gamma$,
or  $\propto t$ in the opposite case. The contribution $F_2C_0 r_0 (s)$ either corresponds to 
$s^{\gamma-2\nu -1}$ and translates to $t^{2\nu - \gamma}$ for $2\nu > \gamma$, or to a constant for $2\nu < \gamma$, and is always subdominant. The contribution of $s_2$ is always 
subdominant as well. 

In conclusion we find that the behavior for short aging times reproduces the behavior of the ordinary walk up to prefactors, as one might expect, and has the same range of convergence, namely, the second
moment exists for $\gamma > 2(\nu-\eta)$.

\section{Long aging times $t_a >> t$ \label{sec:considerably}}

\subsection{Calculation of $F_0$ and $F_2$}

Here again the cases $\gamma<1$ and $\gamma >1$ have to be distinguished. 

\subsubsection{$\gamma<1$}
In this domain we can reuse the previous result in Eq.(\ref{eqn:psi1}), but we now expand it for $t_a>>t$:
\begin{equation}
 F_0(t|t_a) = \frac{\sin \pi \gamma}{\pi} \left( \frac{t_a}{t}\right)^\gamma \frac{1}{t+t_a} 
  \simeq  \frac{\sin \pi \gamma}{\pi} t_a^{\gamma-1} t^{-\gamma},
\end{equation}
so that we get in the Laplace domain 
\begin{equation}
 F_0(s|t_a) \simeq \frac{\sin \pi \gamma}{\pi} \Gamma(1-\gamma) t_a^{\gamma-1} s^{\gamma-1}.
\end{equation}

For $F_{2}(t|t_a)$ we can use our result for $k(t)$, Eq.(\ref{eqn:kGammaSmall}), and insert it into (\ref{eqn:F2}): 
\begin{eqnarray}
F_{2}(t|t_a) &=& \gamma\frac{\sin(\pi \gamma)}{\pi}  c^2 \int^{t_a}_0 (t_a+t-t')^{2(\nu-\eta)} \nonumber \\
&& \;\; \times [(t_a+t-t')^\eta-(t_a-t')^\eta]^2  \nonumber \\
&& \;\;  \times \frac{(t')^{\gamma-1}}{(t_0+t_a+t-t')^{1+\gamma}} dt'.
\end{eqnarray}
Neglecting $t_0$ in the expression in the last line we get
\begin{eqnarray}
&& F_{2}(t|t_a) \simeq  \gamma \frac{\sin(\pi \gamma)}{\pi}  c^2 t_a^{2\nu-2} \int^{t_a}_0 \left(1+\frac{t}{t_a}-\frac{t'}{t_a} \right)^{2(\nu-\eta)-\gamma-1} \nonumber \\
&& \;\; \times \left[ \left(1+\frac{t}{t_a}-\frac{t'}{t_a} \right)^\eta-\left(1- \frac{t'}{t_a}\right)^\eta \right]^2 \left( \frac{t'}{t_a}\right)^{\gamma-1} dt' .
\end{eqnarray}
We introduce the dimensionless variables $z = 1-\frac{t'}{t_a}$ and $y = \frac{t}{t_a}$ and rewrite the integral:
\begin{eqnarray}
&& F_{2}(t|t_a) = \gamma \frac{\sin(\pi \gamma)}{\pi}  c^2 t_a^{2\nu-1} \times \\
&& \;\; \times \int^{1}_0 (z+y)^{2(\nu-\eta)-1-\gamma}[(z+y)^\eta-z^\eta]^2 (1-z)^{\gamma-1} dz . \nonumber
\end{eqnarray}
Since we are going to encounter integrals of this type several times, we will calculate them generally. The general form
\begin{equation}
I_{a,b,c}(y) = \int^{1}_0 (z+y)^{a}[(z+y)^c-z^c]^2 (1-z)^{b} dz \; , \label{eqn:Iabc} 
\end{equation}
can be expressed in terms of Gau{\ss} hypergeometric functions, leading to the following asymptotic behavior for $y \to 0$
\begin{equation}
I_{a,b,c}(y) \simeq \left\{ \begin{array}{l l l}
C(a,c) y^{1+a+2c}    & \mathrm{for} &  a+2c < 1 \\
 \mathrm{B}(1+b,a+2c-1)y^2 c^2  & \mathrm{for} & a+2c > 1.
\end{array} \right. \label{eqn:IabcAsymptotic}
\end{equation}
A detailed derivation and the bounds for the constant $C(a,c)$ are given in Appendix \ref{sec:integral}.

The behavior of $F_{2}$ follows with the substitutions $a= 2(\nu-\eta)-\gamma-1$, $b=\gamma-1$, $c=\eta$. Omitting dimensionless constants we obtain 
two distinct regimes in the limit $t_a>>t$, depending on the relation between $\nu$ and $\gamma$:
\begin{equation}
F_{2}(t|t_a) \propto 
\left\{ \begin{array}{l l}
c^2 t_a^{2\nu-3} t^2 & \text{for } 2\nu> \gamma+2\\
c^2 t_a^{\gamma-1} t^{2\nu-\gamma}   & \text{for } 2\nu < \gamma+2 .
\end{array} \right. 
\end{equation}
Since both of these cases belong to the non-integrable class, we obtain in the Laplace-domain:
\begin{equation}
F_{2}(s|t_a) \propto   \left\{ \begin{array}{l l l}
c^2 t_a^{2\nu-3} s^{-3} & \mathrm{for} & 2\nu> \gamma+2\\
c^2  t_a^{\gamma-1} s^{\gamma-2\nu-1} & \mathrm{for} & 2\nu < \gamma+2 .
\end{array} \right.
\end{equation}

\subsubsection{$\gamma>1$}
By substituting $k(t')=\tau^{-1}$ one finds 
\begin{equation}
 F_0 = \int_0^{t_a} \psi(t_a + t- t') \frac{1}{\tau} dt' \to \frac{1}{\tau} \Psi(t) \simeq \frac{t_0^\gamma}{\tau} t^{-\gamma}.
\end{equation}
Since $\gamma > 1$, the term $F_0$ is of the integrable type, and therefore
\begin{equation}
 F_0(s|t_a) \simeq const. 
\end{equation}
Now we turn to $F_2$.  Starting from equation (\ref{eqn:F2}) one finds 
\begin{eqnarray}
&& F_{2}(t|t_a) \simeq   \gamma c^2 \frac{t_0^{\gamma} }{\tau}  t_a^{2\nu-\gamma-1}\int^{t_a}_0 \left(1+\frac{t}{t_a}-\frac{t'}{t_a} \right)^{2(\nu-\eta)-\gamma-1} \nonumber \\
&& \qquad \times \left[ \left(1+\frac{t}{t_a}-\frac{t'}{t_a} \right)^\eta-\left(1- \frac{t'}{t_a}\right)^\eta \right]^2  dt' \; .
\end{eqnarray}
The calculation is similar to the one in the case $\gamma<1$. With Eq.(\ref{eqn:IabcAsymptotic}) for $I_{2(\nu-\eta)-\gamma-1,0,\eta}(t/t_a)$ we obtain 
\begin{eqnarray}
&& F_{2}(t|t_a) \simeq  \gamma c^2 \frac{ t_0^{\gamma} }{\tau}    \\
&& \;\;\; \times \left\{ \begin{array}{l l l}
\eta^2 B(1, 2\nu -\gamma -2) t_a^{2\nu-\gamma-2}t^2  &\mathrm{for} & 2\nu> \gamma+2\\
 C \cdot t^{2\nu-\gamma} & \mathrm{for} & 2\nu < \gamma+2.
\end{array} \right. \nonumber 
\end{eqnarray}
The case $2\nu> \gamma+2$ still belongs in the non-integrable class and therefore transforms into 
\begin{equation}
F_{2}(s|t_a) \simeq 2 \gamma \eta^2 B(1,2\nu -\gamma -2) c^2 \frac{t_0^{\gamma}}{\tau}   t_a^{2\nu-\gamma-2}  s^{-3} ,
\end{equation}
however for $2\nu < \gamma+2$ we have to distinguish between $\gamma -1 < 2\nu < \gamma+2$, where the $F_2$ is non-integrable, and $2\nu < \gamma -1$, where it is integrable. 
Therefore:
\begin{eqnarray}
&& F_{2}(s|t_a) \simeq \gamma  c^2\frac{ t_0^{\gamma}}{\tau} \\
&& \qquad \times \left\{ \begin{array}{l l l}
\eta^2\mathrm{B}(1,2\nu-\gamma-2) t_a^{2\nu-\gamma-2}  s^{-3} & 2\nu> \gamma+2\\
C \, \Gamma(2\nu+1-\gamma) s^{\gamma-2\nu-1} & \gamma+2> 2\nu > \gamma -1\\
const\,t_0^{2\nu+1-\gamma } & 2\nu < \gamma -1.
\end{array} \right. \nonumber .
\end{eqnarray}

\subsection{Calculation of $s_2$}

\subsubsection{$\gamma<1$}
The calculations for $s_2$ from (\ref{eqn:s2}) are very similar to that for $F_{2}$ case and yield
\begin{eqnarray}
&& s_{2}(t|t_a) \simeq \gamma\frac{ 1}{\gamma-2(\nu-\eta)} \frac{\sin(\pi \gamma)}{\pi} c^2 \\
&& \; \times \left\{ \begin{array}{l l l}
\eta^2 B(2\nu -\gamma -1,\gamma)  t_a^{2\nu-2} t^2 & \mathrm{for}& 2\nu> \gamma+1\\
C \, t_a^{\gamma-1}t^{2\nu + 1-\gamma} & \mathrm{for} & 2\nu < \gamma+1.
\end{array} \right. \nonumber
\end{eqnarray}

\subsubsection{$\gamma>1$}
In this case we have 
\begin{eqnarray}
&& s_{2}(t|t_a) \simeq \gamma \frac{1}{\gamma-2(\nu-\eta)}c^2 \frac{t_0^{\gamma}}{\tau} \nonumber\\
&& \times \int^{t_a}_{0} \left[ (t_a+t-t')^{\eta}-(t_a-t')^{\eta} \right]^2 (t_a+t-t')^{2(\nu-\eta)-\gamma} dt' \nonumber \\
&&  =\gamma \frac{1}{\gamma-2(\nu-\eta)}I_{2(\nu-\eta)-\gamma,0,\eta} c^2 \frac{t_0^{\gamma}}{\tau} (t_a)^{2\nu+1- \gamma} \left(\frac{t}{t_a}\right) .
\end{eqnarray}
Using Eq.($\ref{eqn:IabcAsymptotic}$) again we find
\begin{eqnarray}
&& s_{2}(t|t_a) \simeq \gamma \frac{1}{\gamma-2(\nu-\eta)} c^2 \frac{t_0^{\gamma}}{\tau} \\
&& \;\;\; \times  \left\{ \begin{array}{l l l}
\eta^2  B(2\nu -\gamma -1,1)t_a^{2\nu -\gamma -1} t^2  &\mathrm{for} & 2\nu > \gamma +1 \\
C t^{2\nu +1-\gamma } & \mathrm{for} & 2\nu < \gamma + 1.
\end{array} \right. \nonumber
\end{eqnarray}
The corresponding results for the case of long aging times are summarized in Table \ref{tab:sFstrongAging}.

\subsection{Mean squared displacement}

With these results we can now compute the MSD in the stronlgy aged case. Using Tables \ref{tab:Cr} and \ref{tab:sFstrongAging} 
we can write down the asymptotic behavior of the combinations $F_0 C_0 r_2$, $F_0 C_2 r_0$ and $F_2 C_0 r_0$ in the Laplace domain. 
The inverse transforms are then performed using Tauberian theorems. The corresponsing results for $\gamma < 1$ and for $\gamma > 1$ are summarized in 
Tables \ref{tab:x2StronglyAgedGammaSmall} and \ref{tab:x2StronglyAgedGammaBig}. 

Therefore for considerably aged walks we have:
\subsubsection{$\gamma<1$:}
\begin{equation}
  \langle x^2(t) \rangle \propto  \left\{
  \begin{array}{lll}
   t_a^{\gamma-1} t & \mathrm{for} & 2\nu < \gamma  \\
   t_a^{\gamma-1} t^{2\nu+1-\gamma} & \mathrm{for} &\gamma <2 \nu < 1+\gamma \\ 
  t_a^{2\nu-2} t^2 & \mathrm{for} & 1+\gamma < 2\nu. 
  \end{array}
  \right.
\end{equation}
Obviously, no Richardson's superballistic regime $\langle x^2 \rangle \propto t^3$ is possible in the aged walk. 
This fact by itself does not make the model to be a poor candidate for description of turbulent dispersion, since 
the Richardson's law in turbulence corresponds to starting the observation of the interparticle distance immediately after introducing the tracers
into the stationary flow, i.e. to an ordinary, non-aged case.

\subsubsection{$\gamma> 1$ }
\begin{equation}
  \langle x^2(t) \rangle \propto  \left\{
  \begin{array}{lll}
    t & \mathrm{for} & 2\nu < \gamma  \\
    t^{2\nu+1-\gamma} &\mathrm{for} & \gamma <2 \nu < 1+\gamma \\ 
   t_a^{2\nu-\gamma-1 }t^2 & \mathrm{for} & 1+\gamma < 2\nu.
  \end{array}
  \right.
\end{equation}
Therefore in the regime where the ordinary walk shows normal diffusion or enhanced diffusion there are no or weak changes due to aging (differing only by prefactor). 
In the regime where the ordinary walk is superballistic, it ages to a ballistic one. This finding complies with the fact that a ballistic walk with $\nu = 1$ shows only weak aging, i.e. again the ballistic aged behavior \cite{Froemberg,Magdziarz}. 

\begin{widetext}
\begin{center}
\begin{table}[t!]
 \begin{tabular}{||l|l|l||}
 \hline \hline
\rule[-4mm]{0cm}{1cm}   $\gamma<1$ & $ F_0(s|t_a) \propto t_a^{\gamma-1}s^{\gamma-1}$ & $F_{2}(s|t_a) \propto   c^2 \left\{ \begin{array}{l l}
  t_a^{2\nu-3}  s^{-3} & 2\nu> \gamma+2\\
  t_a^{\gamma-1} s^{\gamma-2\nu-1}  & 2\nu < \gamma+2
\end{array} 
 \right. $\\ \hline
\rule[-4mm]{0cm}{1cm} $\gamma<1$ & $  $ & $s_{2}(t|t_a) \propto  c^2 
\left\{ \begin{array}{l l}
 t_a^{2\nu-2}t^2  & 2\nu> \gamma+1\\
 t_a^{\gamma-1} t^{2\nu+1-\gamma}  & 2\nu < \gamma+1
\end{array} \right. $ \\ \hline
\rule[-4mm]{0cm}{1cm} $\gamma>1$ & $F_0(s|t_a) \propto  1$ & $F_{2}(s|t_a) \propto    c^2  
 \left\{ \begin{array}{l l}
t_0^{\gamma-1}  t_a^{2\nu-\gamma-2} s^{-3} & 2\nu> \gamma+2\\
t_0^{\gamma-1}  s^{\gamma-2\nu-1} & \gamma+2> 2\nu > \gamma -1\\
 \; t_0^{2\nu} & 2\nu < \gamma -1
\end{array} \right.$ \\ \hline
\rule[-4mm]{0cm}{1cm} $\gamma>1$& & $s_{2}(t|t_a) \propto  c^2 \left\{ \begin{array}{l l}
t_0^{\gamma-1}   t_a^{2\nu -\gamma -1}t^2 & 2\nu > \gamma +1 \\
t_0^{\gamma-1}  t^{2\nu +1 -\gamma} & 2\nu < \gamma + 1
\end{array} \right.$ \\ \hline \hline
\end{tabular}
\caption{Results for $F_0$ and $F_2$ in the Laplace domain as well as $s_2$ in the time domain for different parameter ranges in the case of long 
aging times $t_a>>t>>t_0$. Dimensionless prefactors are omitted. \label{tab:sFstrongAging}}
\end{table}
\end{center}

\begin{center}
\begin{table}[h!]
 \begin{tabular}{||l|l|l|l|l||}
 \hline \hline
 \rule[-4mm]{0cm}{1cm}  & $F_2 C_0 r_0 $ & $F_0 C_0 r_2 $ & $F_0 C_2 r_0 $ & $s_2$ \\ \hline
\rule[-4mm]{0cm}{1cm}  $2\nu < \gamma $ & $c^2 t_a^{\gamma-1} t^{2\nu+1-\gamma}$  &  $c^2 t_a^{\gamma-1} t^{2\nu+1-\gamma}$ & \bm{$c^2 t_0^{2\nu-\gamma} t_a^{\gamma-1} t$} & $c^2 t_a^{\gamma-1} t^{2\nu+1-\gamma}$ \\ \hline
\rule[-4mm]{0cm}{1cm} $\gamma < 2\nu < 1+\gamma$ & \bm{$c^2 t_a^{\gamma-1} t^{2\nu+1-\gamma}$} & \bm{$c^2 t_a^{\gamma-1} t^{2\nu+1-\gamma}$} & \bm{$c^2 t_a^{\gamma-1} t^{2\nu+1-\gamma}$} & \bm{$c^2 t_a^{\gamma-1} t^{2\nu+1-\gamma}$} \\ \hline
\rule[-4mm]{0cm}{1cm} $1+\gamma < 2\nu < 2 + \gamma$ & $c^2 t_a^{\gamma-1} t^{2\nu+1-\gamma}$  & $c^2 t_a^{\gamma-1} t^{2\nu+1-\gamma}$ & $c^2 t_a^{\gamma-1} t^{2\nu+1-\gamma}$ & \bm{$c^2 t_a^{2\nu-2} t^2$} \\ \hline
\rule[-4mm]{0cm}{1cm} $2+\gamma < 2\nu$  & $c^2 t_a^{2\nu-3}t^3$  & $c^2 t_a^{\gamma-1} t^{2\nu+1-\gamma}$ &  $c^2 t_a^{\gamma-1} t^{2\nu+1-\gamma}$ & \bm{$c^2 t_a^{2\nu-2} t^2$} \\ \hline \hline
\end{tabular}
\caption{Asymptotic behavior of the contributions to the MSD for $\gamma<1$ in the limit $t_a>>t>>t_0$. All dimensionless prefactors are omitted. The dominant terms are highlighted in boldface.
\label{tab:x2StronglyAgedGammaSmall}}
\end{table}

\begin{table}[h!]
 \begin{tabular}{||l|l|l|l|l||}
 \hline \hline
 \rule[-4mm]{0cm}{1cm}  & $F_2 C_0 r_0 $ & $F_0 C_0 r_2 $ & $F_0 C_2 r_0 $ & $s_2$ \\ \hline
\rule[-4mm]{0cm}{1cm}  $2\nu < \gamma-1$ & $c^2 t_0^{2\nu}  $ & $c^2 t_0^{2\nu } $ & \bm{$c^2 t_0^{2\nu-1}  t$}  & $c^2 t_0^{\gamma-1}  t^{2\nu+1-\gamma}$ \\ \hline
\rule[-4mm]{0cm}{1cm}  $ \gamma-1 <2\nu < \gamma $ & $c^2 t_0^{\gamma-1}  t^{2\nu+1-\gamma}$ & $c^2 t_0^{\gamma-1}  t^{2\nu+1-\gamma}$  & \bm{$c^2 t_0^{2\nu -1} t $} & $c^2 t_0^{\gamma-1}  t^{2\nu+1-\gamma}$ \\ \hline
\rule[-4mm]{0cm}{1cm} $\gamma < 2\nu < 1+\gamma$ & \bm{$c^2 t_0^{\gamma-1} t^{2\nu+1-\gamma}$}   & \bm{$c^2 t_0^{\gamma-1} t^{2\nu+1-\gamma}$}   & \bm{$c^2 t_0^{\gamma-1} t^{2\nu+1-\gamma}$}   & \bm{$c^2 t_0^{\gamma-1} t^{2\nu+1-\gamma}$}  \\ \hline
\rule[-4mm]{0cm}{1cm} $1+\gamma< 2\nu <2 + \gamma$  & $c^2 t_0^{\gamma-1} t^{2\nu+1-\gamma}$  &  $c^2 t_0^{\gamma-1} t^{2\nu+1-\gamma}$  & $c^2 t_0^{\gamma-1} t^{2\nu+1-\gamma}$   & \bm{$c^2 t_0^{\gamma-1} t_a^{2\nu-\gamma-1 }t^2$} \\ \hline 
\rule[-4mm]{0cm}{1cm} $2+\gamma<2\nu$  & $c^2 t_0^{\gamma-1} t_a^{2\nu-\gamma-2} t^{3}$ & $c^2 t_0^{\gamma-1} t^{2\nu+1-\gamma}$   & $c^2 t_0^{\gamma-1} t^{2\nu+1-\gamma}$   & \bm{$c^2 t_0^{\gamma-1} t_a^{2\nu-\gamma-1 }t^2$} \\ \hline \hline
\end{tabular}
\caption{Asymptotic behavior of the contributions to the MSD for $\gamma>1$ in the limit $t_a>>t>>t_0$. All dimensionless prefactors are omitted. The dominant terms are highlighted in boldface. \label{tab:x2StronglyAgedGammaBig}}
\end{table}
\end{center}
\end{widetext}

Summarizing the findings for the case of considerably aged walks we state that for the parameter range where the ordinary walk showed subdiffusion we now find 
regular diffusion but with a prefactor that decays with growing aging times. 
There is no place for the Richardson's regime in the considerably aged case.
Again, the second moment extsts only for $\gamma>2(\nu-\eta)$, and if it exists the value of $\eta$ only enters the prefactors, but does not change the power-law
dependences of the MSD on all times involved. 

\begin{figure}[H]
\begin{center}
\scalebox{0.48}{\includegraphics{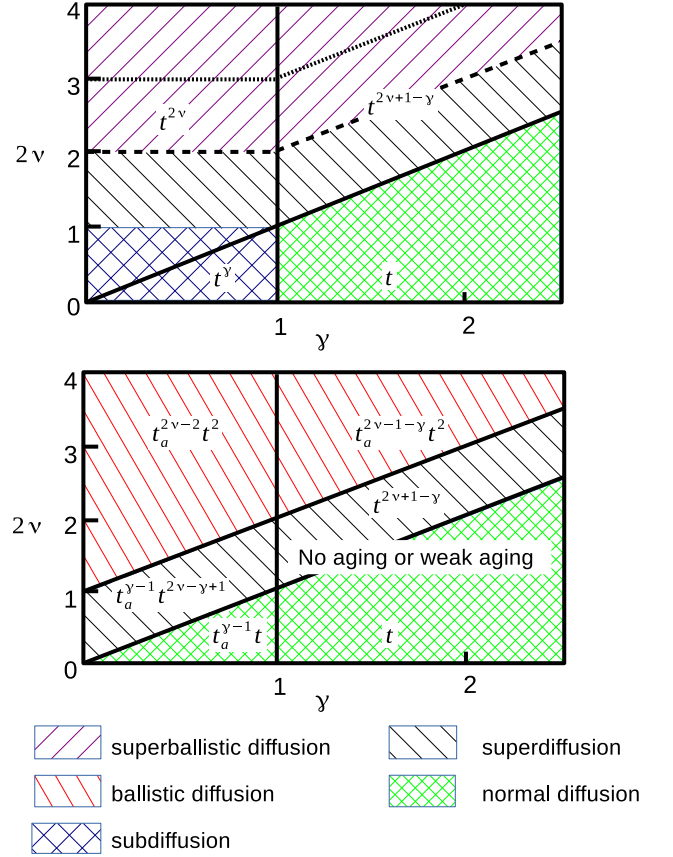}}
\caption{The upper panel shows the map of behaviors of the ensemble average $\langle x^2(t) \rangle$ in the ordinary Drude walk. The thick solid lines correspond to the
changes in time-dependences while the hatchings represent the type of diffusion. The dashed line corresponds to the ballistic behavior, and the dotted one to the Richardson's law.
The lower panel shows the same for the considerably aged walk. 
\label{MSDPlot} }
\end{center}
\end{figure}

We moreover note that the double time-ensemble average $\langle \langle x^2 (t) \rangle_T \rangle_E$, also discussed in Ref. \cite{Radons}, whose calculation involves 
an additional integration over the time,$\langle \langle x^2 (t) \rangle_T \rangle_E \simeq (T-t)^{-1} \int_0^{T-t} \langle x^2 (t|t_a) \rangle dt_a$  shows the same behavior as 
the aged walk if the measurement time $t$ is associated with the time lag in the double average, 
and the aging time $t_a$ is changed for the data acquisition time $T$.   

\section{Conclusions \label{Sec:Conclusions}}

We considered a generalized L\'evy walk model interpolating between the original L\'evy walk model of \cite{Shlesinger} (characterized by two parameters $\gamma$ and $\nu$)
and Drude-like models of \cite{Schulz,BarKlaf}, as proposed in the Supplemental material of Ref. \cite{Radons}. The model contains an additional parameter $\eta$ which allows for such an interpolation: 
the case $\eta = 1$ corresponds to the original L\'evy walk, the case $\eta = \nu$ to a Drude-like model. We show that the mean squared displacement in such a walk 
is finite for the sets of parameters obeying $\gamma > 2(\nu-\eta)$ and diverges otherwise. 
The temporal evolution of the MSD follows the same patterns for all values of $\eta$ for which it does not diverge; the particular value of $\eta$ enters only the prefactors.  
In particular, Drude models never show divergences, and possess the Richardson's regime in the ordinary case. 

While ordinary walks show a plentitude of regimes ranging from subdiffusive to superballistic behavior, the considerably aged walks only show the behavior ranging from
diffusive to ballistic. The results of the calculations are best summed up in Fig. \ref{MSDPlot}. 

\section{Acknowledgements} The authors are thankful to Dr. M.F. Shlesinger for turning their attention to the problem. 
Partial financial support by DFG within the IRTG 1740 research and training group (Mercator fellowship of F. Sagues) is gratefuly acknowledged. 

\appendix

\section{A note on Tauberian theorems} \label{sec:tauberian}

In the present work we continuously employed direct are inverse Laplace transforms of functions asymptotically following power laws. 
The asymptotic forms of Laplace transforms of functions which at long times behave as $f(t) = t^{\rho-1} L(t)$, where $L(t)$ is a slowly varying function, 
in the asymptotic regime $s \to 0$, and back transforms may be obtained by use of Tauberian theorems. 
We assume that a Laplace transform $f(s) = \int_0^t f(t) e^{-st} dt$ exists, i.e. the function $f(t)$ does not possess a strong divergence at 0. 
All functions $f(t)$ appearing in our work are non-negative. 
For such functions Laplace transforms are monotonically decaying functions of $s$.
Depending on the behavior of $f(t)$ at infinity two cases should be considered.  

The function $f(t)$ might be integrable on $[0, \infty)$, so that 
$\int_0^\infty f(t) dt = I_0^{f} < \infty$, or this integral may diverge. The first case corresponds to $\rho < 0$ and 
the second one to $\rho > 0$ (the case $\rho = 0$ may belong to the either class depending on the concrete form of $L(t)$). 

In the second case a Tauberian theorem may be applied immediately, stating that if $f(t)$ is a regularly varying function, i.e. when  its Laplace transform is given by 
\begin{equation}
 f(t) \simeq t^{\rho-1} L(t) \;\; \leftrightarrow \;\; f(s) \simeq \Gamma(\rho) s^{-\rho} L\left(\frac{1}{s}\right)
 \label{eq:Tauberian}
\end{equation}
for $\rho \geq 0$. As in the main text, all slowly varying functions will be omitted (i.e. changed for constants $L$). 

We note that for $\rho < 0$ Eq.(\ref{eq:Tauberian}) gives $f(s)$ being a growing function of $s$ and therefore is wrong. In this case let us consider the function
\begin{equation}
 S(t) = \int_t^\infty f(t') dt'.
\end{equation}
The integrability of $f(t)$ means that $S(t)$ is well-defined, and that $I_0^{f} =  \int_0^\infty f(t') dt' = S(0)$ is finite. 
The function $S(t)$ has the power-law asymptotics 
\begin{equation}
 S(t) \simeq \frac{L t^\rho}{\rho},
\end{equation}
and, if this is no more integrable (i.e. for $\rho > -1$), can be transformed via the Tauberian theorem, so that
\begin{equation}
 S(s) \simeq L \frac{\Gamma(\rho+1)}{\rho} s^{-(\rho +1)} = L \Gamma(\rho)  s^{-(\rho +1)},
\end{equation}
where in the last equality the identity $\Gamma(x+1) = x \Gamma(x)$ was used. 
Noting that $f(t) = - \frac{d}{dt}S(t)$ and using the Laplace representation of the derivative, we get 
\begin{equation}
 f(s) = S(t=0) - s S(s) = I^{f}_0 - L \Gamma(\rho)  s^{-\rho} .
\end{equation}
The direct application of the Tauberian theorem would give us a correct form of the singular term (up to a sign), but not the regular one. 

If $S(t)$ is still integrable, we consider the function $P(t) = \int_t^\infty S(t') dt'$, whose power-law asymptotics for $t \to \infty$ is
\begin{equation}
 P(t) \simeq \frac{L t^{\rho+1}}{\rho (\rho+1)},
\end{equation}
and whose connection to $f(t)$ is given by $f(t) = \frac{d^2}{dt^2}S(t)$. For $-2 < \rho$ the function $P(t)$ is not integrable, 
and the application of the Tauberian theorem gives
\begin{equation}
 P(s) = L \frac{\Gamma(\rho+2)}{\rho (\rho+1)} s^{-\rho-2} = L \Gamma(\rho) s^{-\rho-2}.
\end{equation}
Using the Laplace representation for the second derivative we get
\begin{equation}
 f(s) = - s P(t=0) - P'(t=0) + s^2 P(s). 
\end{equation}
The value of $P'(t=0)$ is $-S(t=0)=-I^{f}_0$. The value $P(t=0)$ is given by the integral
\begin{equation}
P(t=0) = \int_0^\infty dt \int_t^\infty f(t')dt'.
\end{equation}
Changing the sequence of integrations in $t$ and $t'$ we get 
\begin{equation}
 P(t=0) = \int_0^\infty dt' f(t') \int_0^{t'} dt = \int_0^\infty t' f(t') dt'.
\end{equation}
Since $f(t)$ decays with $t$ faster than $t^{-2}$, the integral converges, and will be denoted by $I^{f}_1$. 
Therefore we have
\begin{equation}
 f(s) = I^{f}_0 - s I^{f}_1 + L \Gamma(\rho) s^{-\rho}.
\end{equation}

For $\rho < -2$ the procedure has to be repeated again for the function being the integral of $P(t)$, etc. The general result is 
\begin{align}
 f(s) = \sum_{k=0}^{k_{\max}} \frac{(-1)^k}{k!} I^{f}_k s^k +(-1)^{k_{\max}+1} L \Gamma(\rho) s^{-\rho}
\end{align}
with $k_{\max}$ being the whole part of $-\rho$, and $I^{f}_k$ being the moment integral
\begin{align}
I^{f}_k = \int_0^\infty t^k f(t) dt.
\end{align}
In the main text we never have to use more than first three terms of this expansion.

\section{Estimates for the integral $I_{a,b,c}(y)$ \label{sec:integral}} 

We are interested in the integral 
\begin{eqnarray}
I_{a,b,c}(y) &=& \int^{1}_0 (1-z)^{b} [(z+y)^c-z^c]^2 (z+y)^{a}  dz \label{eqn:Iabc1} \\ 
&=& \int^{1}_0 (1-z)^{b} \left[ (z+y)^{a+2c} -2 (z+y)^{a+c} z^{c} \right. \nonumber \\
&&  \left. + (z+y)^{a} z^{2c}\right] dz \nonumber 
\end{eqnarray}
in the limit of small $y= \frac{t}{t_a} \ll 1$ for the parameter ranges $c > 0$, $b > -1$, $a \in \mathbb{R}$.

To evaluate it we use the Euler's integral representation for the Gau{\ss} hypergeometric function \cite{AbraSteg} for $\Re \; c' > \Re \; b' > 0$
\begin{eqnarray}
_{2}F_{1}(a',b';c';x) &=& \frac{1}{\mathrm{B}(b',c'-b')} \times  \label{IntegralRep} \\
&& \times \int_{0}^{1} z^{b'-1} (1-z)^{c'-b'-1} (1-zx)^{-a'}. \nonumber
\end{eqnarray}
As the existence condition $1+b>0$ is always satisfied for all three terms in (\ref{eqn:Iabc1}) we can write the integral as
\begin{eqnarray}
&& I_{a,b,c}(y) = \label{eqn:Iabc2} \\
&& y^a  \left[  y^{2c} \mathrm{B}(1,1+b) _2F_1 \left(-a-2c,1;2+b; -\frac{1}{y} \right) \right. \nonumber \\
&& -2 y^{c} \mathrm{B}(1+c, 1+b) _2F_1 \left(-a-c,1+c;2+b+c; -\frac{1}{y} \right) \nonumber \\ 
&& \left. + \mathrm{B}(1+2c , 1+b) _2F_1 \left(-a,1+2c;2+b+2c; -\frac{1}{y} \right) \right] dz  \nonumber 
\end{eqnarray}
with $\mathrm{B}(x,y)$ being the Beta function. 
Athough the integral can be expressed in terms of three Gau{\ss} hypergeometric functions,
its investigation is somewhat tricky, since the asymptotic regimes appear as a subleading terms 
in a sum of three large contributions whose leading terms cancel. First, to avoid evaluating hypergeometric functions at $-\infty$ 
we make use of the Pfaff transformations:
\begin{align}
_2F_1(a',b';c';z) = (1-z)^{-b'} \; _2F_1 \left( b',c'-a';c;\frac{z}{z-1} \right) \label{Pfaff1} \\
_2F_1(a',b';c';z) = (1-z)^{-a'} \; _2F_1 \left( a',c'-b';c;\frac{z}{z-1} \right). \label{Pfaff2}
\end{align}
These two different forms will be useful in different domains of parameters. Under the transformations the argument of the corresponding functions on the r.h.s., equal to $\frac{1}{1+y}$, will tend to 1.
%We essentially do not need the Pfaff transformations and simply make the corresponding changes of variables and regrouping in the initial form of the integral,
%but these essentially are quite obscure. 

Applying the Pfaff transformation Eq.(\ref{Pfaff1}) to the integrals in Eq.(\ref{eqn:Iabc2}) we find:
\begin{eqnarray*}
&& I_{a,b,c}(y) = y^{1+a+2c} \times \\
&& \times \bigg[  (1+y)^{-1} \mathrm{B}(1,1+b) \times  \\
&& \qquad  \times \;  _2F_1 \left(1,2+a+b+2c;2+b; \frac{1}{1+y} \right)  \\ 
&&  -2 (1+y)^{-1-c} \mathrm{B}(1+c, 1+b) \times  \\
&& \qquad  \times \;  _2F_1 \left(1+c,2+a+b+2c;2+b+c; \frac{1}{1+y} \right) \\ 
&&  + (1+y)^{-1-2c} \mathrm{B}(1+2c , 1+b) \times  \\
&& \qquad \left. \times \;  _2F_1 \left(1+2c,2+a+b+2c;2+b+2c;\frac{1}{1+y} \right) \right].
\end{eqnarray*} 
We now use the Euler integral representation (\ref{IntegralRep}) again, but switch between $a'$ and $b'$:
\begin{eqnarray*}
_{2}F_{1}(a',b';c';x) &= &\frac{1}{\mathrm{B}(a',c'-a')} \times \\
&& \times \int_{0}^{1} z^{a'-1} (1-z)^{c'-a'-1} (1-zx)^{-b'} 
\end{eqnarray*}
for $\Re \; c' > \Re \;  a' > 0$.
Note that the existence condition for the integrals is the same as before, $b+1>0$, which is satisfied for all relevant cases in this paper, so we can write:
\begin{eqnarray*}
&& I_{a,b,c}(y) = y^{1+a+2c} \times  \\
&& \times \int_{0}^{1}  \left[  (1+y)^{-1}  (1-z)^{b} \left( 1- \frac{z}{1+y} \right)^{-2-a-b-2c}  \right. \\ 
&& \qquad \left. -2 (1+y)^{-1-c} z^c (1-z)^{b} \left( 1- \frac{z}{1+y} \right)^{-2-a-b-2c} \right. \\ 
&& \qquad \left. + (1+y)^{-1-2c} z^{2c} (1-z)^{b} \left( 1- \frac{z}{1+y} \right)^{-2-a-b-2c} \right] dz .
\end{eqnarray*}
The integrals of each of three contributions in square brackets would diverge for $y \to 0$, but the integral of whole sum is convergent for $a+2c < 1$ since for $y \to 0$ the integrand tends to 
\[
(1-2 z^{c} + z^{2c}) (1-z)^{-2-a-2c} =  (1-z^c)^2 (1-z)^{-2-a-2c},
\]
and the integral
\[
 C(a,c) = \int_0^1 (1-z^c)^2 (1-z)^{-2-a-2c} dz
\]
of this expression converges in the range $a+2c <1$ (to prove the convergence it is enough to expand the first term in vicinity of $z=1$). 
This integral cannot be expressed in terms of ``simple'' functions, but the (loose) bounds for it follow easily. 

Let us find two constants $B > A > 0$ such that for all $0< z < 1$
\[
 A (1-z) < 1-z^c < B (1-z). 
\]
To do so consider the function
\[
 f(z)=\frac{1-z^c}{1-z},
\]
with $f(0)=1$ and with its limiting value at $z \to 1$ given by the l'H\^opital's rule $\lim_{z \to 1} = c$. 
Therefore the limit of the function at 1 is larger than its value at $0$ when $c>1$ and smaller than this value when $c<1$. For $c=1$ this function equals to unity identically.

Now we consider $c \neq 1$ and proseed to show that the 
function $f(z)$ is monotonically growing for $c>1$ and monotonically decaying for $c<1$. To show this it is enough to show that its derivative 
on $[0,1]$ does not vanish. The derivative of the corresponding function is 
\[
 f'(z)=\frac{1-z^c+cz^c-cz^{c-1}}{(1-z)^2},
\]
and can only vanish when the numerator, $g(z)= 1-z^c+cz^c-cz^{c-1}$, vanishes somewhere at $0\leq z < 1$. Vanishing of the numerator at $z=1$ does not pose a problem since $f'(z)$ diverges and tends to $(c-1)(1-z)^{-2}$ for $z=1$, 
being positive in vicinity of $z=1$ for $c>1$ and negative for for $c < 1$ due to the fact that the denominator vanishes even faster.
Now we show that this function never changes its sign on $0\leq z < 1$.
Calculating the derivative 
\[
 g'(z)=- c(c-1)z^{c-2} + c(c-1)z^{c-1} = - c(c-1)z^{c-2}(1-z)
\]
we see that it is strictly positive for all $z<1$ for $c<1$ and strictly negative for $c>1$. Therefore the bounds for the function $f(z)$ are 
given by its limiting values of 1 at $z=0$ and $c$ at $z = 1$. Therefore we have $A=\min(1,c^2)$ and $B=\max(1,c^2)$. 
Since for $1>a+2c $
\[
 \int_0^1 (1-z)^{-a-2c} dz = \frac{1}{1-a-2c}
\]
we get 
\begin{equation}
\frac{\min(1,c^2)}{1-a-2c} \leq C \leq \frac{\max(1,c^2)}{1-a-2c}. \label{eq:bounds}
\end{equation}
Therefore, for  $a+2c < 1$ we have, for $y$ small, 
\begin{equation}
I_{a,b,c }(y) \simeq C y^{1+a+2c}
\label{eq:I1A}
\end{equation}  
where the bounds for the constant $C$ are given by Eq.(\ref{eq:bounds}).

For the opposite case $a+2c>1$ we have to use the other Pfaff transformation, Eq.(\ref{Pfaff2}), resulting in:
\begin{eqnarray}
&& I_{a,b,c}(y) = (1+y)^{a} \times \\
&& \times \big[  (1+y)^{2c} \mathrm{B}(1,1+b) \times \nonumber \\
&& \quad \times  _2F_1 \left(-a-2c,1+b;2+b; 1/(1+y) \right) \nonumber \\ 
&& -2 (1+y)^{c} \mathrm{B}(1+c, 1+b) \times \nonumber \\
&& \quad \times  _2F_1 \left(-a-c,1+b;2+b+c; 1/(1+y) \right)  \nonumber \\ 
&& +  \mathrm{B}(1+2c , 1+b) \times \nonumber \\
&& \quad \times _2F_1 \left(-a,1+b;2+b+2c;1/(1+y) \right) \big]. \nonumber 
\end{eqnarray} 

We now use the integral representation Eq.(\ref{IntegralRep}) again and find 
\begin{eqnarray}
&& I_{a,b,c}(y) = (1+y)^{a} \int_0^1 \left[  (1+y)^{2c} z^b \left( 1-\frac{z}{1+y} \right)^{a+2c}   \right. \nonumber\\
&& \qquad -2 (1+y)^{c} z^b (1-z)^{c} \left( 1-\frac{z}{1+y} \right)^{a+c} \nonumber  \\ 
&& \qquad \left. + z^b (1-z)^{2c} \left( 1-\frac{z}{1+y} \right)^{a}  \right] dz. \label{eqn:Iabc3}
\end{eqnarray} 
Now we expand the expression in each term of the integrand up to the second order in $y$ using the fact that 
\[
(1+y)^{\alpha} \simeq 1 + \alpha y + \frac{\alpha(\alpha-1)}{2} y^2
\]
and 
\begin{eqnarray*}
&& \left( 1-\frac{z}{1+y} \right)^{\alpha} \simeq (1-z)^{\alpha} +  \alpha (1-z)^{\alpha-1} z y \\
&& \;\;\; + \frac{1}{2} \left[ \alpha (\alpha-1) (1-z)^{\alpha-2} z^2 -2 \alpha (1-z)^{\alpha-1}z \right] y^2,
\end{eqnarray*}
and using the definition of the Beta function
\[
\mathrm{B}(a,b) = \int^1_0 z^{a-1} (1-z)^{b-1} dz
\]
we find:
\begin{eqnarray*}
&& I_{a,b,c}(y) \simeq c^2 y^2 [ \mathrm{B}(1+b,1+a+2c) \\
&& \qquad +2\mathrm{B}(2+b,a+2c) + \mathrm{B}(3+b,a+2c-1)].
\end{eqnarray*}
The first two orders in $y$ have canceled, so the leading term goes as $y^2$. From the argument of the last Beta function it is also clear, that the result only holds for $a+2c>1$, 
i.e. exactly in the parameter range where Eq.(\ref{eq:I1A}) ceases to be applicable, and that the exponents are continuous at $a+2c=1$. 
Rewriting the Beta functions as $\mathrm{B}(a,b) = \Gamma(a) \Gamma(b)/ \Gamma(a+b)$ and (repeatedly) using the identity $\Gamma(x+1) = x \Gamma(x)$ we find a compact representation
of the sum of the three beta functions, namely 
\begin{equation}
 I_{a,b,c}(y) \simeq c^2 y^2 \mathrm{B}(1+b,a+2c-1).
\end{equation}
In conclusion we have:
\[
I_{a,b,c}(y) \simeq \left\{ \begin{array}{l l l}
C(a,c) y^{1+a+2c} & \mathrm{for} & a+2c< 1 \\
 c^2 \mathrm{B}(1+b,a+2c-1) y^2  & \mathrm{for} & a+2c> 1, 
\end{array} \right.
\]
which is the Eq.(\ref{eqn:IabcAsymptotic}) of the main text, with the bounds on a constant $C(a,c)$ given by Eq.(\ref{eq:bounds}).


\begin{thebibliography}{99}
\bibitem{Shlesinger} M.F. Shlesinger, B.J. West, and J. Klafter, Phys. Rev. Lett. \textbf{58}, 110 (1987)
\bibitem{Zaburdaev} V. Zaburdaev, S. Denisov, and J. Klafter, Rev. Mod. Phys. \textbf{87}, 483 (2015) 
\bibitem{Pearson} K. Pearson, Nature \textbf{72}, 294 (1905)
\bibitem{Swinney} T.H. Solomon, E.R. Weeks, and H.L. Swinney, Phys. Rev, Lett \textbf{71}, 3975 (1993)
\bibitem{Swinney2} T.H. Solomon, E.R. Weeks, and H.L. Swinney, Physica D \textbf{76}, 70 (1994)
\bibitem{Poeschke} P. P\"oschke, I.M. Sokolov, M.A. Zaks, and A.A. Nepomnyashchy, Phys. Rev. E \textbf{96}, 062128 (2017)
\bibitem{Richardson} L.F. Richardson, Proc. Roy. Soc. London \textbf{110}, 709 (1926)
\bibitem{Shlesinger2} J. Klafter, A. Blumen, and M. F. Shlesinger, Phys. Rev. A \textbf{35}, 3081 (1987)
\bibitem{Radons} T. Albers and G. Radons, Phys. Rev. Lett. \textbf{120}, 104501 (2018)
\bibitem{Drude} Drude P., Ann. Phys. (Leipzig), \textbf{1}, 566 (1900)
\bibitem{Schulz} H. Schulz-Baldes, Phys. Rev. Lett. \textbf{78}, 2176 (1997) 
\bibitem{BarKlaf} E. Barkai and J. Klafter, in: S. Benkadda and G.M. Zaslavsky, eds., \textit{Chaos, Kinetics and Nonlinear Dynamics in Fluids and Plasmas} (Lecture Notes in Physics v. 511) Springer,
Berlin Heidelberg, 1998
\bibitem{Sokolov_Drude1} I.M. Sokolov, A. Blumen, and J. Klafter, EPL \textbf{47}, 152 (1999)
Ballistic versus diffusive pair dispersion in the Richardson regime
\bibitem{Sokolov_Drude2} I.M. Sokolov, J. Klafter, and A. Blumen, Phys. Rev. E \textbf{61}, 2717 (2000)
\bibitem{Boffetta} G. Boffetta and I.M. Sokolov, Physics of Fluids \textbf{14}, 3224 (2002)
\bibitem{BryPr} A.P. Prudnikov, Yu.A. Brychkov and O.I. Marichev, \textit{Integrals and Series} V. 1, Gordon and Breach, NY, 1986
\bibitem{Sokolov_Klafter} I.M. Sokolov and J. Klafter, \textit{First Steps in Random walks: From Tools to Applications}, Oxford University press, Oxford, 2011
\bibitem{Magdziarz} M. Magdziarz and T. Zorawik, Phys. Rev. E \textbf{95}, 022126 (2017)
\bibitem{Froemberg} D. Froemberg and E. Barkai, Phys. Rev. E \textbf{87}, 030104(R) (2013)
\bibitem{AbraSteg} M. Abramovitz and I.A.Stegun, eds., Handbook of mathematical functions, Dover, N.Y. 1972, Sec. 15. 
\end{thebibliography}
\end{document}